\documentclass[letter,10pt]{article}

\usepackage{times}
\usepackage{graphicx}
\usepackage{subfigure}
\usepackage{epstopdf}
\usepackage{amsfonts}
\usepackage{amssymb}
\usepackage{amsmath}
\usepackage{makecell}
\usepackage{dsfont}
\usepackage{threeparttable}
\usepackage{pifont}
\usepackage{array}
\usepackage{multirow}
\usepackage{cite}
\usepackage{caption}
\usepackage{endnotes}
\usepackage{booktabs}

\newcommand{\mb}[1]{\mathbf{#1}}
\newcommand{\mc}[1]{\mathcal{#1}}
\newtheorem{theorem}{Theorem}[section]
\newtheorem{property}{Property}[section]

\begin{document}

\title{STAIR Codes: A General Family of Erasure Codes for Tolerating Device
	and Sector Failures\footnote{An earlier version of this work was presented at the 12th USENIX Conference on
File and Storage Technologies (FAST '14), Santa Clara, CA, February, 2014
\cite{Li14}.  This extended version includes new reliability analysis on STAIR
codes, and discusses how to configure the sector failure coverage of STAIR
codes (see \S\ref{sec:reliability}).}}

\author{Mingqiang Li and Patrick P. C. Lee\\
\emph{Department of Computer Science and Engineering, The Chinese University of Hong Kong}\\
\emph{Email: mingqiangli.cn@gmail.com, pclee@cse.cuhk.edu.hk}}

\maketitle

\begin{abstract}
Practical storage systems often adopt erasure codes to tolerate device
failures and sector failures, both of which are prevalent in the field.
However, traditional erasure codes employ device-level redundancy to protect
against sector failures, and hence incur significant space overhead.
Recent sector-disk (SD) codes are available only for limited configurations.
By making a relaxed but practical assumption, we construct a general family of
erasure codes called {\em STAIR codes}, which efficiently and provably
tolerate both device and sector failures without any restriction on the size
of a storage array and the numbers of tolerable device failures and sector
failures.  We propose the \emph{upstairs encoding} and
\emph{downstairs encoding} methods, which
provide complementary performance advantages for different configurations.  We
conduct extensive experiments on STAIR codes in
terms of space saving, encoding/decoding speed, and update cost.  We
demonstrate that STAIR codes not only improve space efficiency over
traditional erasure codes, but also provide better computational efficiency
than SD codes based on our special code construction.  Finally, we present
analytical models that characterize the reliability of STAIR codes, and show
that the support of a wider range of configurations by STAIR codes is critical
for tolerating sector failure bursts discovered in the field.
\end{abstract}

\section{Introduction}
\label{sec:intro}


Mainstream disk drives are known to be susceptible to both
{\em device failures} \cite{Pinheiro07,Schroeder07} and
{\em sector failures} \cite{Bairavasundaram07,Schroeder10}: a device failure
implies the loss of all data in the failed device, while a sector failure
implies the data loss in a particular disk sector.  In particular, sector
failures are of practical concern not only in disk drives, but also in
emerging solid-state drives (SSDs) as they often appear as worn-out blocks after
frequent program/erase cycles \cite{Grupp09,Boboila10,Grupp12,Zheng13}.
In the face of device and sector failures, practical storage systems often
adopt {\em erasure codes} to provide data redundancy \cite{Plank13a}.
However, existing erasure codes often build on tolerating device failures and
provide device-level redundancy only.  To tolerate additional sector failures,
an erasure code must be constructed with extra parity
disks.  A representative example is RAID-6,
which uses two parity disks to tolerate one device failure together with one
sector failure in another non-failed device \cite{White10,Intel05}.
If the sector failures can span a number of devices, the same number of
parity disks must be provisioned.  Clearly, dedicating an entire parity disk
for tolerating a sector failure is too extravagant.

To tolerate both device and sector failures in a space-efficient
manner, sector-disk (SD) codes \cite{Plank13b,Plank13c} and the earlier PMDS
codes \cite{Blaum13a} (which are a subset of SD codes) have recently been
proposed.
Their idea is to introduce parity sectors, instead of entire parity disks, to
tolerate a given number of sector failures.  However, the constructions of
SD codes are known only for limited configurations (e.g., the number of
tolerable sector failures is no more than three), and some of the known
constructions rely on exhaustive searches \cite{Plank13b,Plank13c,Blaum13c}.
An open issue is to provide a general construction of erasure codes that can
efficiently tolerate both device and sector failures without any restriction
on the size of a storage array, the number of tolerable device failures, or
the number of tolerable sector failures.

In this paper, we make the first attempt to develop such a generalization,
which we believe is of great theoretical and practical interest to provide
space-efficient fault tolerance for today's storage systems.  After carefully
examining the assumption of SD codes on failure coverage, we find that
although SD codes have relaxed the assumption of the earlier PMDS codes to
comply with how most storage systems really fail, the assumption
remains too strict.  By reasonably relaxing the assumption of SD codes on
sector failure coverage, we construct a general family of erasure codes called
\emph{STAIR codes}, which efficiently tolerate both device and sector
failures.

Specifically, SD codes devote $s$ sectors per stripe to coding, and tolerate
the failure of any $s$ sectors per stripe.  We relax this assumption in STAIR
codes by limiting the number of devices that may simultaneously contain sector
failures, and by limiting the number of simultaneous sector failures per
device.
Consequently, as shown in
\S\ref{sec:problem}, STAIR codes are constructed to protect the sector failure
coverage defined by a vector $\mb{e}$, rather than all combinations of $s$
sector failures.


With the relaxed assumption, the construction of STAIR codes can be based on
existing erasure codes.  For example, STAIR codes can build on Reed-Solomon
codes (including standard Reed-Solomon codes \cite{Reed60,Plank97,Plank05} and
Cauchy Reed-Solomon codes \cite{Blomer95,Plank06}), which have no restriction
on code length and fault tolerance.

We first define some basic concepts and elaborate how the sector failure
coverage is formulated for STAIR codes in \S\ref{sec:problem}.  Then the paper
makes the following contributions:
\begin{itemize}
\item
We present a baseline construction of STAIR codes. Its idea is to run
two orthogonal encoding phases based on Reed-Solomon codes.
See \S\ref{sec:construction}.
\item
We propose an {\em upstairs decoding} method, which systematically
reconstructs the lost data due to both device and sector failures. The proof
of fault tolerance of STAIR codes follows immediately from the decoding
method. See \S\ref{sec:recovery}.
\item
Inspired by upstairs decoding, we extend the construction of STAIR codes to
regularize the code structure.  We propose two encoding
methods: {\em upstairs encoding} and {\em downstairs encoding}, both of which
reuse computed parity results in subsequent encoding.  The two encoding
methods provide complementary performance advantages for different
configuration parameters.  See \S\ref{sec:transformation}.
\item
We extensively evaluate STAIR codes in terms of space saving, encoding/decoding
speed, and update cost.  We show that STAIR codes achieve significantly higher
encoding/decoding speed than SD codes through parity reuse.  Most importantly,
we show the versatility of STAIR codes in supporting any size of a storage
array, any number of tolerable device failures, and any number of tolerable
sector failures.  See \S\ref{sec:evaluation}.
\item
We develop analytical models to characterize the reliability of STAIR codes 
and discuss how the sector failure coverage of STAIR codes should be
configured.  We examine both independent and correlated sector failure models, 
and show that it is critical for STAIR codes to support a wider range of
configurations in the presence of sector failure bursts discovered in the
field \cite{Bairavasundaram07,Schroeder10}. See \S\ref{sec:reliability}.
\end{itemize}

We review related work in \S\ref{sec:related}, and conclude this paper in
\S\ref{sec:conclusions}.

\section{Preliminaries}
\label{sec:problem}

This section presents the definitions and the problem of simultaneous device
and sector failures in storage arrays.  Table~\ref{tab:parameter-stair}
summarizes the major notation used for the STAIR code construction. 

\begin{table}[t]
  \centering
  \caption{Major notation used for the STAIR code construction.}
  \label{tab:parameter-stair}
  \begin{small}
  \begin{tabular}{lp{4.5in}}
    \toprule
    \textbf{Notation} & \hfil \textbf{Description} \hfil \\
    \midrule
	\multicolumn{2}{l}{Defined in \S\ref{sec:problem}:}\\
    $n$ & Number of chunks per stripe (i.e. number of devices per storage array)\\
    $r$ & Number of sectors (i.e. symbols) per chunk\\
    $m$ & Maximum number of entirely failed chunks (due to device failures) per stripe\\
	$m'$ & Maximum number of partially failed chunks (due to sector failures) per stripe\\
	$\mb{e}$ & Sector failure coverage, defined as $\mb{e}=(e_0,e_1,\cdots,e_{m'-1})$  (where $0 < e_0 \le e_1 \le \cdots \le e_{m'-1} \leq r$)\\
    $s$ & Maximum number of sector failures per stripe, defined as $s = \sum_{i=0}^{m'-1} e_i$\\
	&\\
	\multicolumn{2}{l}{Defined in \S\ref{sec:construction}:}\\
	$d_{i,j}$ & Data symbol (where $0\le i\le r-1$, and $0\le j\le n-m-1$)\\
	$p_{i,k}$ & Row parity symbol (where $0\le i\le r-1$, and $0\le k\le m-1$)\\
	$p'_{i,l}$ & Intermediate parity symbol (where $0\le i\le r-1$, and $0\le l\le m'-1$)\\
	$g_{h,l}$ & Outside global parity symbol (where $0\le l\le m'-1$, and $0\le h\le e_l-1$)\\
    $\mc{C}_{row}$ & Systematic MDS code for encoding in row direction\\
    $\mc{C}_{col}$ & Systematic MDS code for encoding in column direction\\
	&\\
	\multicolumn{2}{l}{Defined in \S\ref{sec:recovery}:}\\
    $d^*_{h,j}$ & Virtual parity symbol encoded from a data chunk (where $0\le h\le e_l-1$, and $0\le j\le n-m-1$)\\
	$p^*_{h,k}$ & Virtual parity symbol encoded from a row parity chunk (where $0\le h\le e_l-1$, and $0\le k\le m-1$)\\
	&\\
	\multicolumn{2}{l}{Defined in \S\ref{sec:transformation}:}\\
    $\hat{g}_{h,l}$ & Inside global parity symbol (where $0\le l\le m'-1$, and $0\le h\le e_l-1$)\\
    \bottomrule
  \end{tabular}
  \end{small}
\end{table}

\begin{figure}[t]
\centering
\includegraphics[width=3.5in]{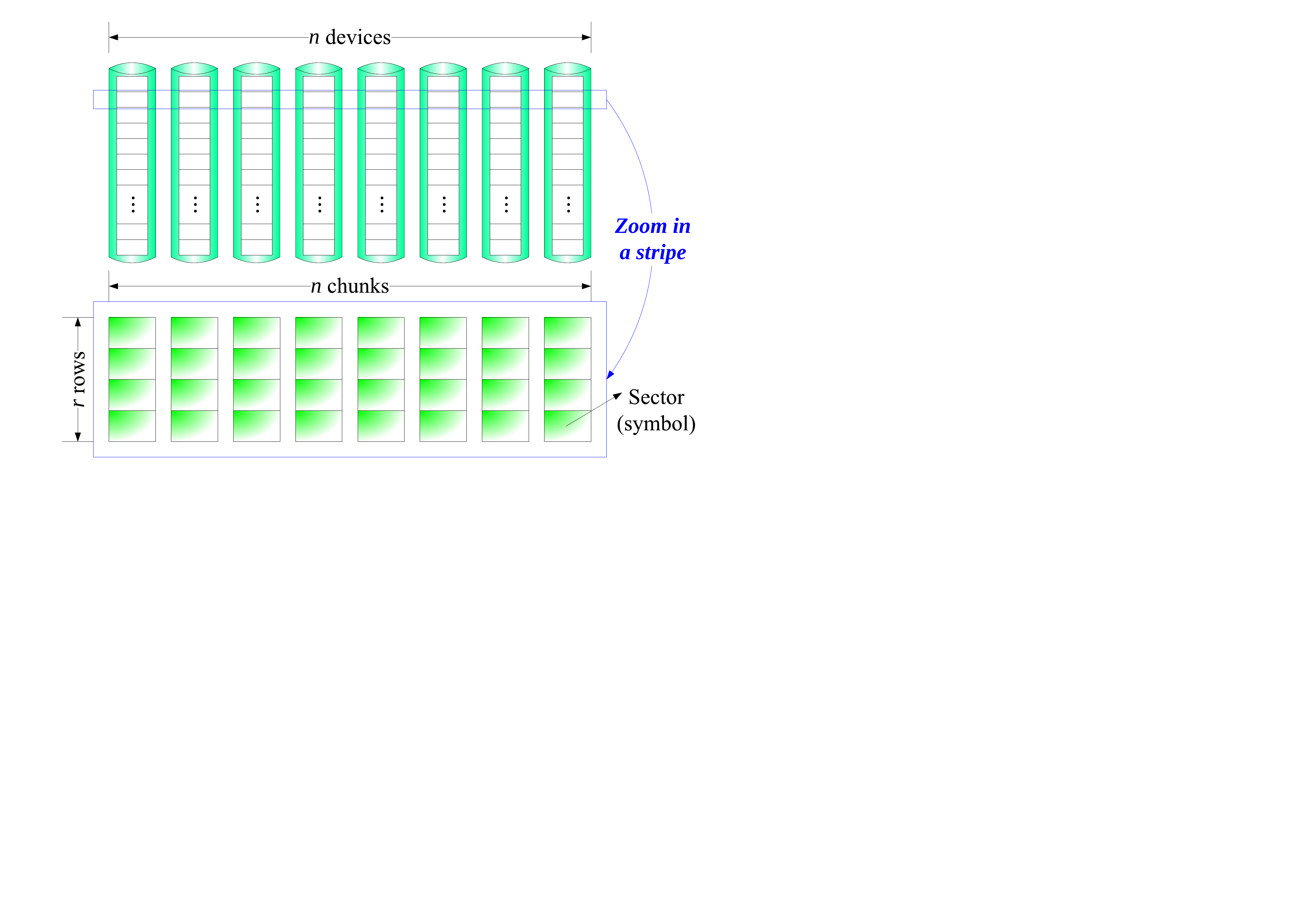}
\caption{A stripe for $n=8$ and $r=4$.}
\label{fig:stripe}
\end{figure}

We consider a storage array with $n$ devices, each of which has its storage
space logically segmented into a sequence of continuous
\emph{chunks} (also called \emph{strips})
of the same size. We group each of the $n$ chunks at the same position
of each device into a \emph{stripe}, as depicted in Figure~\ref{fig:stripe}.
Each chunk is composed of $r$ sectors.  Thus, we can view the
stripe as a $r\times n$ array of sectors.  Using coding theory
terminology, we refer to each sector as a \emph{symbol}. Each stripe
is independently protected by an erasure code for fault tolerance, so our
discussion focuses on a single stripe.

Storage arrays are subject to both device and sector failures.  A device
failure can be mapped to the failure of an entire chunk of a stripe.  We
assume that the stripe can tolerate at most $m$ ($< n$) chunk failures, in
which all symbols are lost.  In addition to device failures, we assume that
sector failures can occur in the remaining $n-m$ devices. Each sector
failure is mapped to a lost symbol in the stripe.
Suppose that besides the $m$ failed chunks, the stripe can tolerate sector
failures in at most $m'$ ($\le n-m$) remaining chunks, each of which has
a maximum number of
sector failures defined by a vector $\mb{e} =(e_0, e_1, \cdots, e_{m'-1})$.
Without loss of generality, we arrange the elements of $\mb{e}$ in
monotonically increasing order (i.e., $e_0 \le e_1 \le \cdots \le e_{m'-1}$).
For example, suppose that sector failures can only simultaneously appear in at
most three chunks (i.e., $m'=3$), among
which at most one chunk has two sector failures and the remaining have one
sector failure each.  Then, we can express $\mb{e} = (1,1,2)$.  Also, let $s =
\sum_{i=0}^{m'-1} e_i$ be the total number of sector failures defined by
$\mb{e}$.  Our study assumes that the configuration parameters $n$, $r$, $m$,
and $\mb{e}$ (which then determines $m'$ and $s$) are the inputs selected by
system practitioners for the erasure code construction.

Erasure codes have been used by practical storage systems to protect
against data loss \cite{Plank13a}.  We focus on a class of erasure codes with
optimal storage efficiency called {\em maximum distance separable (MDS)}
codes, which are defined by two parameters $\eta$ and $\kappa$ ($< \eta$).  We
define an $(\eta,\kappa)$-code as
an MDS code that transforms $\kappa$ symbols into $\eta$ symbols collectively
called a {\em codeword} (this operation is called {\em encoding}), such that
any $\kappa$ of the $\eta$ symbols can be used to recover the original $\kappa$
uncoded symbols (this operation is called {\em decoding}).
Each codeword is encoded from $\kappa$ uncoded symbols by multiplying
a row vector of the $\kappa$ uncoded symbols with a $\kappa\times\eta$
{\em generator matrix} of coefficients based on Galois Field arithmetic.  We
assume that the $(\eta,\kappa)$-code is {\em systematic}, meaning that the
$\kappa$ uncoded symbols are kept in the codeword.  We refer to the $\kappa$
uncoded symbols as {\em data} symbols, and the $\eta-\kappa$ coded symbols as
{\em parity} symbols.  We use systematic MDS codes as the building blocks of
STAIR codes.  Examples of such codes are standard Reed-Solomon codes
\cite{Reed60,Plank97,Plank05} and Cauchy Reed-Solomon codes
\cite{Blomer95,Plank06}.

Given parameters $n$, $r$, $m$, and $\mb{e}$ (and
hence $m'$ and $s$), our goal is to construct a STAIR code that tolerates
both $m$ failed chunks and $s$ sector failures in the remaining $n-m$ chunks
defined by $\mb{e}$.  Note that some special cases of $\mb{e}$ have the
following physical meanings:
\begin{itemize}
\item If $\mb{e} = (1)$, the corresponding STAIR code is equivalent to a
PMDS/SD code with $s=1$ \cite{Blaum13a,Plank13b,Plank13c}. In fact, the STAIR
code is a new construction of such a PMDS/SD code.
\item If $\mb{e} = (r)$, the corresponding STAIR code has the same
function as a systematic $(n,n-m-1)$-code.
\item If $\mb{e} = (\epsilon,\epsilon,\cdots,\epsilon)$ with $m'=n-m$ and some
constant $\epsilon<r$, the corresponding STAIR code has the same function
as an intra-device redundancy (IDR) scheme
\cite{Dholakia08,Dholakia11,Schroeder10} that adopts a systematic
$(r,r-\epsilon)$-code.
\end{itemize}

We show via examples how we can define the sector failure coverage vector
$\mb{e}$ in STAIR codes in practice.  We provide more formal analysis on the
configurations of $\mb{e}$ in \S\ref{sec:reliability}. 

We argue that STAIR codes can be configured to provide more general protection
than SD codes \cite{Plank13b,Plank13c,Blaum13c}.  One major use case of STAIR
codes is to protect against bursts of contiguous sector failures
\cite{Bairavasundaram07,Schroeder10}.  Let $\beta$ be the maximum length of a
tolerable sector failure burst in a chunk.  Then we should set $\mb{e}$ with
its largest element $e_{m'-1}=\beta$.  For example, when $\beta=2$, we may set
$\mb{e}$ as our previous example $\mb{e} = (1,1,2)$, or a weaker and
lower-cost $\mb{e} = (1,2)$.  In some extreme cases, some disk models may have
longer sector failure bursts (e.g., with $\beta > 3$) \cite{Schroeder10}.
Take $\beta=4$ for example. Then we can define $\mb{e} = (1,4)$, so that the
corresponding STAIR code can tolerate a burst of four sector failures in one
chunk together with an additional sector failure in another chunk.  In
contrast, such an extreme case cannot be handled by SD codes, whose current
construction can only tolerate at most three sector failures in a stripe
\cite{Plank13b,Plank13c,Blaum13c}.  Thus, although the numbers of device and
sector failures (i.e., $m$ and $s$, respectively) are often small in practice,
STAIR codes support a more general coverage of device and
sector failures, especially for extreme cases.

STAIR codes also provide more space-efficient protection than the IDR scheme
\cite{Dholakia08,Dholakia11,Schroeder10}.  To protect against a burst of
$\beta$ sector failures in {\em any} data chunk of a stripe, the IDR scheme
requires $\beta$ additional redundant sectors in each of the $n-m$
data chunks.  This is equivalent to setting $\mb{e} = (\beta, \beta, \cdots,
\beta)$ with $m'=n-m$ in STAIR codes.  In contrast, the general construction
of STAIR codes allows a more flexible definition of $\mb{e}$, where $m'$ can
be less than $n-m$, and all elements of $\mb{e}$ except the largest element
$e_{m'-1}$ can be less than $\beta$.  For example, to protect against a burst
of $\beta=4$ sector failures for $n=8$ and $m=2$ (i.e., a RAID-6
system with eight devices), the IDR scheme introduces a total of $4\times6 =
24$ redundant sectors per stripe; if we define $\mb{e} = (1,4)$ in STAIR codes
as above, then we only introduce five redundant sectors per stripe.
Thus, STAIR codes introduce fewer redundant sectors than the IDR scheme in
general.

\section{Baseline Encoding}
\label{sec:construction}


\begin{figure}[t]
\centering
\includegraphics[width=4.5in]{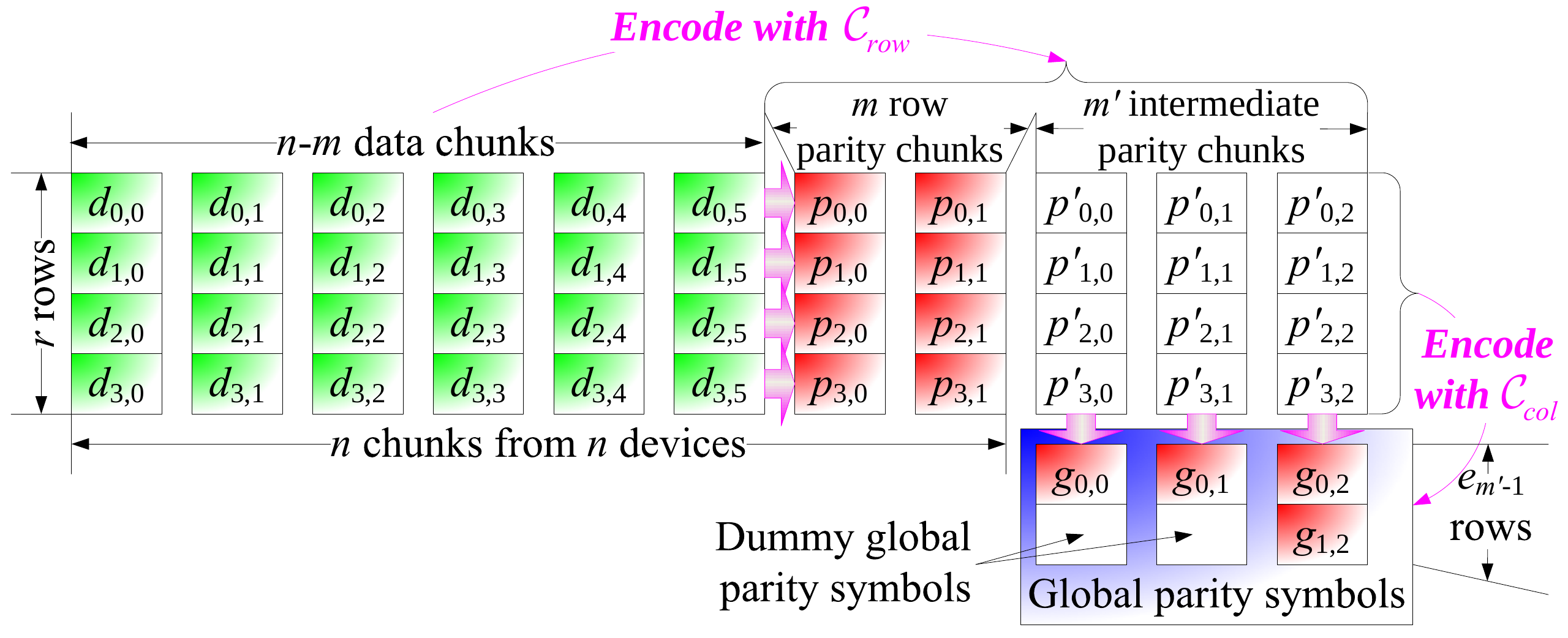}
\caption{Exemplary configuration: a STAIR code stripe for $n=8$, $r=4$, $m=2$,
and $\mb{e}=(1, 1, 2)$ (i.e., $m'=3$ and $s=4$). Throughout this paper, we use
this configuration to explain the operations of STAIR codes.}
\label{fig:construction}
\end{figure}

For general configuration parameters $n$, $r$, $m$, and $\mb{e}$, the main
idea of STAIR encoding is to run two orthogonal encoding phases using two
systematic MDS codes.  First, we encode the data symbols using one code and
obtain two types of parity symbols: {\em row parity symbols}, which protect
against device failures, and {\em intermediate parity symbols}, which will
then be encoded using another code to obtain {\em global parity symbols},
which protect against sector failures.  In the following, we elaborate the
encoding of STAIR codes and justify our naming convention.
		
We label different types of symbols for STAIR codes as follows.
Figure~\ref{fig:construction} shows the layout of an exemplary stripe of a
STAIR code for $n=8$, $r=4$, $m=2$, and $\mb{e}=(1, 1, 2)$ (i.e., $m'=3$ and
$s=4$).  A stripe is composed of $n-m$ data chunks and $m$ row parity chunks.
We also assume that there are $m'$ intermediate parity chunks and $s$ global
parity symbols outside the stripe.  Let $d_{i,j}$, $p_{i,k}$, $p'_{i,l}$, and
$g_{h,l}$ denote a data symbol, a row parity symbol, an intermediate parity
symbol, and a global parity symbol, respectively, where $0\le i\le r-1$,
$0\le j\le n-m-1$, $0\le k\le m-1$, $0\le l\le m'-1$, and $0\le h\le e_l-1$.

Figure~\ref{fig:construction} depicts the steps of the two orthogonal
encoding phases of STAIR codes.  In the first encoding phase, we use an
$(n+m',n-m)$-code denoted by $\mc{C}_{row}$ (which is an (11,6)-code in
Figure~\ref{fig:construction}).  We encode via $\mc{C}_{row}$ each row of
$n-m$ data symbols to obtain $m$ row parity symbols and $m'$
intermediate parity symbols in the same row:

\smallskip\noindent
\textbf{Phase 1:} For $i=0,1,\cdots,r-1$,
\begin{equation}\label{equ:phase1}
    d_{i,0},d_{i,1},\cdots,d_{i,n-m-1}  \overset{\mc{C}_{row}}{\Longrightarrow} p_{i,0},p_{i,1},\cdots,p_{i,m-1}, p'_{i,0},p'_{i,1},\cdots,p'_{i,m'-1},
\end{equation}
where $\overset{\mc{C}}{\Longrightarrow}$ describes that the input symbols on
the left are used to generate the output symbols on the right using some code
$\mc{C}$.
We call each $p_{i,k}$ a ``row'' parity symbol since it is only encoded from
the same row of data symbols in the stripe, and we call each $p'_{i,l}$ an
``intermediate'' parity symbol since it is not actually stored but is used in
the second encoding phase only.

In the second encoding phase, we use a $(r+e_{m'-1},r)$-code denoted
by $\mc{C}_{col}$ (which is a (6,4)-code in Figure~\ref{fig:construction}).
We encode via $\mc{C}_{col}$ each chunk of $r$ intermediate parity
symbols to obtain at most $e_{m'-1}$ global parity symbols:

\smallskip\noindent
{\bf Phase 2:} For $l=0,1,\cdots,m'-1$,
\begin{equation}\label{equ:phase2}
p'_{0,l},p'_{1,l},\cdots,p'_{r-1,l}
\overset{\mc{C}_{col}}{\Longrightarrow}
\overset{e_{m'-1}}{\overbrace{g_{0,l},g_{1,l},\cdots,g_{e_l-1,l},*,\cdots,*}},
\end{equation}
where ``$*$'' represents a ``dummy'' global parity symbol that will not be
generated when $e_l < e_{m'-1}$, and we only need to compute the ``real''
global parity symbols $g_{0,l},g_{1,l},\cdots,g_{e_l-1,l}$.
The intermediate parity symbols will be discarded after this encoding
phase.  Note that each $g_{h,l}$ is in essence encoded from all the data
symbols in the stripe, and thus we call it a ``global'' parity symbol.

We point out that $\mc{C}_{row}$ and $\mc{C}_{col}$ can be any systematic MDS
codes.  In this work, we implement both $\mc{C}_{row}$ and $\mc{C}_{col}$
using Cauchy Reed-Solomon codes \cite{Blomer95,Plank06}, which have no
restriction on code length and fault tolerance.

From Figure~\ref{fig:construction}, we see that the logical layout of global
parity symbols looks like a stair. This is why we name this family of erasure
codes \emph{STAIR codes}.

In the following discussion, we use the exemplary configuration in
Figure~\ref{fig:construction} to explain the detailed operations of STAIR
codes.  To simplify our discussion, we first assume that the global parity
symbols are kept outside a stripe and are always available for ensuring
fault tolerance.  In \S\ref{sec:transformation}, we will extend the encoding
of STAIR codes when the global parity symbols are kept inside the stripe and
are subject to both device and sector failures.

\section{Upstairs Decoding}
\label{sec:recovery}

In this section, we justify the fault tolerance of STAIR codes defined by
$m$ and $\mb{e}$. We introduce an {\em upstairs decoding} method
that systematically recovers the lost symbols when both device and sector
failures occur.

\subsection{Homomorphic Property}
\label{subsec:homomorphism}

\begin{figure}[t]
\centering
\includegraphics[width=4.3in]{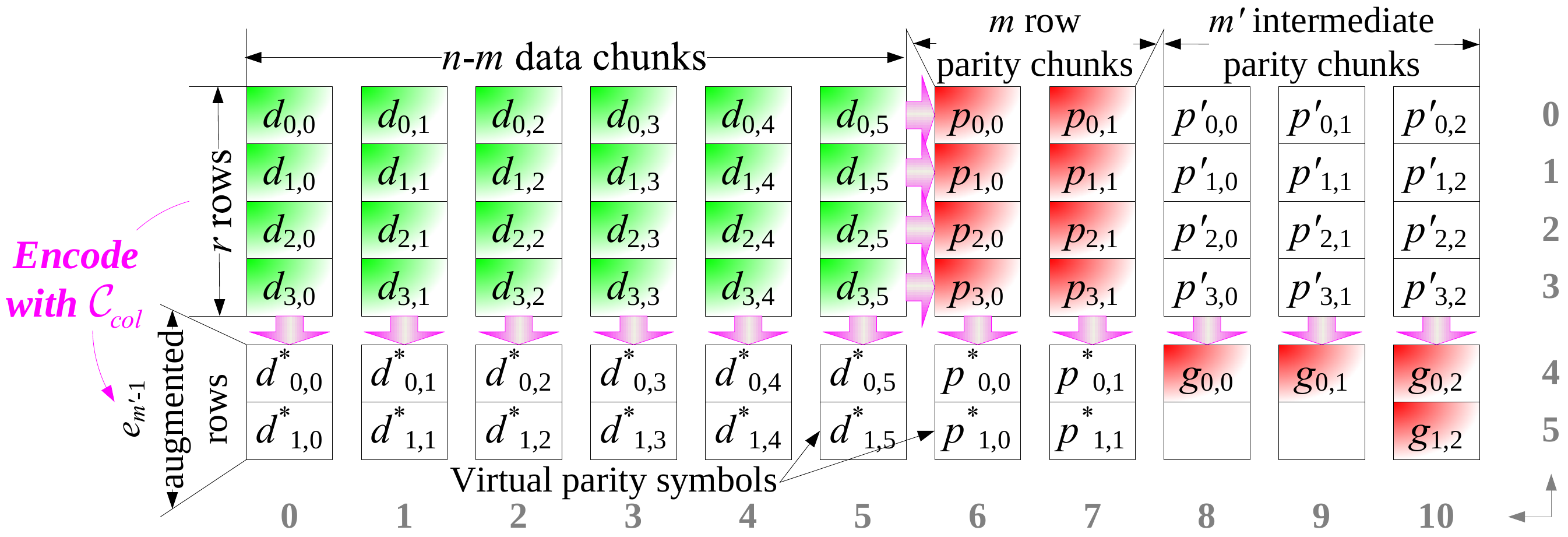}
\caption{A canonical stripe augmented from the stripe in
Figure~\ref{fig:construction}. The rows and columns are labeled from 0 to 5
and 0 to 10, respectively, for ease of presentation.}
\label{fig:canonical}
\end{figure}

The proof of fault tolerance of STAIR codes builds on the concept of a
{\em canonical stripe}, which is constructed by augmenting the existing stripe
with additional {\em virtual parity symbols}.  To illustrate,
Figure~\ref{fig:canonical} depicts how we augment the stripe of
Figure~\ref{fig:construction} into a canonical stripe.  Let $d^*_{h,j}$ and
$p^*_{h,k}$ denote the virtual parity symbols encoded with $\mc{C}_{col}$ from
a data chunk and a row parity chunk, respectively, where $0\le j\le n-m-1$,
$0\le k\le m-1$, and $0\le h\le e_{m'-1}-1$.
Specifically, we use $\mc{C}_{col}$ to generate virtual parity symbols from
the data and row parity chunks as follows:

\smallskip\noindent
For $j=0,1,\cdots,n-m-1$,
\begin{equation}\label{equ:virtualdata}
    d_{0,j},d_{1,j},\cdots,d_{r-1,j} \overset{\mc{C}_{col}}{\Longrightarrow} d^*_{0,j},d^*_{1,j},\cdots,d^*_{e_{m'-1}-1,j};
\end{equation}
and for $k=0,1,\cdots,m-1$,
\begin{equation}\label{equ:virtualparity}
    p_{0,k},p_{1,k},\cdots,p_{r-1,k} \overset{\mc{C}_{col}}{\Longrightarrow} p^*_{0,k},p^*_{1,k},\cdots,p^*_{e_{m'-1}-1,k}.
\end{equation}
The virtual parity symbols $d^*_{h,j}$'s and $p^*_{h,k}$'s, along with the
real and dummy global parity symbols, form $e_{m'-1}$ augmented rows of $n+m'$
symbols. In fact, the resulting canonical stripe in Figure~\ref{fig:canonical}
is a codeword of the product code \cite{Elias54} of $\mc{C}_{row}$ and
$\mc{C}_{col}$.  To make our discussion simpler, we number the rows and
columns of the canonical stripe from 0 to $r+e_{m'-1}-1$ and from 0 to
$n+m'-1$, respectively, as shown in Figure~\ref{fig:canonical}.


Referring to Figure~\ref{fig:canonical}, we know that the upper $r$ rows of
$n+m'$ symbols are codewords of $\mc{C}_{row}$.  We argue that each of the
lower $e_{m'-1}$ augmented rows is in fact also a codeword of $\mc{C}_{row}$.
We call this the {\em homomorphic property}, since the encoding of each chunk
in the column direction preserves the coding structure in the row direction.
We formally prove the homomorphic property in Appendix~\ref{sec:appendix-Homomorphic}.
We use this property to prove the fault tolerance of STAIR codes.

\subsection{Proof of Fault Tolerance}
\label{subsec:recovery_proof}

\begin{figure}[t]
\centering
\includegraphics[width=4in]{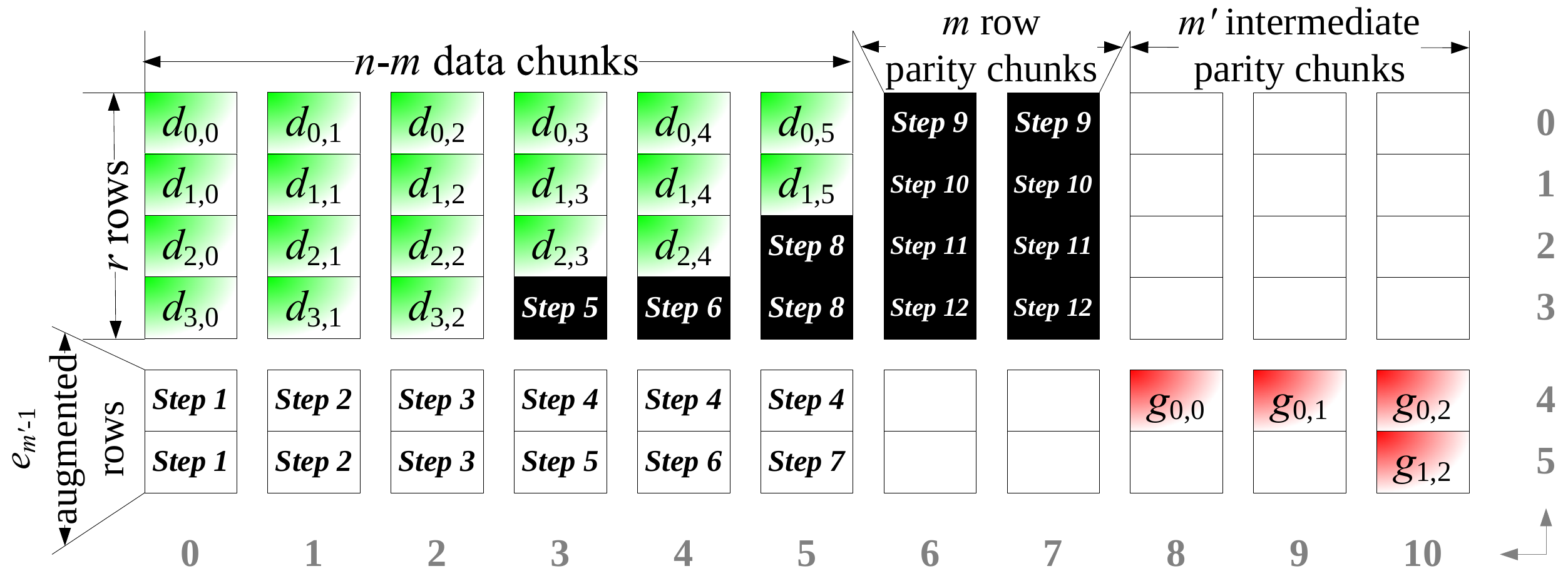}
\caption{Upstairs decoding based on the canonical stripe in
Figure~\ref{fig:canonical}.}
\label{fig:recover_step}
\end{figure}

We prove that for a STAIR code with configuration parameters $n$, $r$, $m$,
and $\mb{e}$, as long as the failure pattern is within the failure coverage
defined by $m$ and $\mb{e}$, the corresponding lost symbols can always be
recovered (or decoded).  In addition, we present an {\em upstairs decoding}
method, which systematically recovers the lost symbols for STAIR codes.

For a stripe of the STAIR code, we consider the worst-case recoverable failure
scenario where there are $m$ failed chunks (due to device failures) and $m'$
additional chunks that have $e_0$, $e_1$, $\cdots$, $e_{m'-1}$ lost
symbols (due to sector failures), where $0 < e_0 \leq e_1 \leq
\cdots \leq e_{m'-1}$.  We prove that all the $m'$ chunks with sector failures
can be recovered with global parity symbols.  In particular, we show that these
$m'$ chunks can be recovered in the order of $e_0$, $e_1$,
$\cdots$, $e_{m'-1}$.  Finally, the $m$ failed chunks due to device failures
can be recovered with row parity chunks.

\subsubsection{Example}

We demonstrate via our exemplary configuration how we recover the lost data
due to both device and sector failures.
Figure~\ref{fig:recover_step} shows the sequence of our decoding steps.
Without loss of generality, we logically assign the column identities such
that the $m'$ chunks with sector failures are in Columns~$n-m-m'$ to
$n-m-1$, with $e_0$, $e_1$, $\cdots$, $e_{m'-1}$ lost symbols, respectively,
and the $m$ failed chunks are in Columns~$n-m$ to $n-1$.  Also, the sector
failures all occur in the bottom of the data chunks.  Thus, the lost symbols
form a stair, as shown in Figure~\ref{fig:recover_step}.

The main idea of upstairs decoding is to recover the lost symbols from left to
right and bottom to top.  First, we see that there are $n-m-m'=3$ good chunks
(i.e., Columns~0-2) without any sector failure.  We encode via
$\mc{C}_{col}$ (which is a (6,4)-code) each such good chunk to obtain
$e_{m'-1}=2$ virtual parity symbols (Steps~1-3).  In Row~4, there are
now six available symbols. Thus, all the unavailable symbols in this row can
be recovered using $\mc{C}_{row}$ (which is a (11,6)-code) due to the
homomorphic property (Step 4).   Note that we only need to recover the $m'=3$
symbols that will later be used to recover sector failures.  Column~3 (with
$e_0=1$ sector failure) now has four available symbols.  Thus, we
can recover one lost symbol and one virtual parity symbol using $\mc{C}_{col}$
(Step 5).  Similarly, we repeat the decoding for Column~4 (with $e_1=1$ sector
failure) (Step 6).  We see that Row~5 now contains six available symbols, so we
can recover one unavailable virtual parity symbol (Step 7).  Then Column~5 (with
$e_2=2$ sector failures) now has four available symbols, so we can recover two
lost symbols (Step 8).  Now all chunks with sector failures are recovered.
Finally, we recover the $m=2$ lost chunks row by row using $\mc{C}_{row}$
(Steps 9-12).  Table~\ref{tab:recovery} lists the detailed decoding steps of
our example in Figure~\ref{fig:recover_step}.

\begin{table}[t]
  \centering
  \caption{Upstairs decoding: detailed steps for the example in Figure~\ref{fig:recover_step}. }
  \label{tab:recovery}

  \begin{tabular}{|c|c@{\ }c@{\ }c|c|}
  \hline
  \textbf{Step} & \multicolumn{3}{c|}{\textbf{Detailed Description}} & \textbf{Coding Scheme} \\
  \hline
  \hline
  1 & $d_{0,0}, d_{1,0}, d_{2,0}, d_{3,0}$ & $\Rightarrow$ & $d^*_{0,0}, d^*_{1,0}$ & $\mc{C}_{col}$ \\
  \hline
  2 & $d_{0,1}, d_{1,1}, d_{2,1}, d_{3,1}$ & $\Rightarrow$ & $d^*_{0,1}, d^*_{1,1}$ & $\mc{C}_{col}$ \\
  \hline
  3 & $d_{0,2}, d_{1,2}, d_{2,2}, d_{3,2}$ & $\Rightarrow$ & $d^*_{0,2}, d^*_{1,2}$ & $\mc{C}_{col}$ \\
  \hline
  4 & $d^*_{0,0}, d^*_{0,1}, d^*_{0,2}, g_{0,0}, g_{0,1}, g_{0,2}$ & $\Rightarrow$ & $d^*_{0,3}, d^*_{0,4}, d^*_{0,5}$ & $\mc{C}_{row}$ \\
  \hline
  5 & $d_{0,3}, d_{1,3}, d_{2,3}, d^*_{0,3}$ & $\Rightarrow$ & $d_{3,3}, d^*_{1,3}$ & $\mc{C}_{col}$ \\
  \hline
  6 & $d_{0,4}, d_{1,4}, d_{2,4}, d^*_{0,4}$ & $\Rightarrow$ & $d_{3,4}, d^*_{1,4}$ & $\mc{C}_{col}$ \\
  \hline
  7 & $d^*_{1,0}, d^*_{1,1}, d^*_{1,2}, d^*_{1,3}, d^*_{1,4}, g_{1,2}$ & $\Rightarrow$ & $d^*_{1,5}$ & $\mc{C}_{row}$ \\
  \hline
  8 & $d_{0,5}, d_{1,5}, d^*_{0,5}, d^*_{1,5}$ & $\Rightarrow$ & $d_{2,5}, d_{3,5}$ & $\mc{C}_{col}$ \\
  \hline
  9 & $d_{0,0}, d_{0,1}, d_{0,2}, d_{0,3}, d_{0,4}, d_{0,5}$ & $\Rightarrow$ & $p_{0,1}, p_{0,2}$ & $\mc{C}_{row}$ \\
  \hline
  10 & $d_{1,0}, d_{1,1}, d_{1,2}, d_{1,3}, d_{1,4}, d_{1,5}$ & $\Rightarrow$ & $p_{1,1}, p_{1,2}$ & $\mc{C}_{row}$ \\
  \hline
  11 & $d_{2,0}, d_{2,1}, d_{2,2}, d_{2,3}, d_{2,4}, d_{2,5}$ & $\Rightarrow$ & $p_{2,1}, p_{2,2}$ & $\mc{C}_{row}$ \\
  \hline
  12 & $d_{3,0}, d_{3,1}, d_{3,2}, d_{3,3}, d_{3,4}, d_{3,5}$ & $\Rightarrow$ & $p_{3,1}, p_{3,2}$ & $\mc{C}_{row}$ \\
  \hline
\end{tabular}
\end{table}


\subsubsection{General Case}
\label{subsec:general}

We now generalize the steps of upstairs decoding.

\textbf{(1) \emph{Decoding of the chunk with $e_0$ sector failures:}} It is
clear that there are $n-(m+m')$ good chunks without any sector failure in the
stripe.  We use $\mc{C}_{col}$ to encode each such good chunk to obtain
$e_{m'-1}$ virtual parity symbols.  Then each of the first $e_0$ augmented rows must
now have $n-m$ available symbols: $n-(m+m')$ virtual parity symbols that have just
been encoded and $m'$ global parity symbols.   Since an augmented row is a
codeword of $\mc{C}_{row}$ due to the homomorphic property, all the
unavailable symbols in this row can be recovered using $\mc{C}_{row}$.
Then, for the column with $e_0$ sector failures, it now has $r$ available
symbols: $r-e_0$ good symbols and $e_0$ virtual parity symbols that have just been
recovered. Thus, we can recover the $e_0$ sector failures as well as the
$e_{m'-1}-e_0$ unavailable virtual parity symbols using $\mc{C}_{col}$.

\textbf{(2) \emph{Decoding of the chunk with $e_i$ sector failures ($1 \leq i
\leq m'-1$):}}   If $e_i=e_{i-1}$, we repeat the decoding for the
chunk with $e_{i-1}$ sector failures.  Otherwise, if $e_i>e_{i-1}$, each of the
next $e_i-e_{i-1}$ augmented rows now has $n-m$ available symbols:
$n-(m+m')$ virtual parity symbols that are first recovered from the good
chunks, $i$ virtual parity symbols that are recovered while the sector
failures are recovered, and $m'-i$ global parity symbols. Thus, all the
unavailable virtual parity symbols in these $e_i-e_{i-1}$ augmented rows can
be recovered.
Then the column with $e_i$ sector failures now has $r$ available symbols:
$r-e_i$ good symbols and $e_i$ virtual parity symbols that have been recovered.
This column can then be recovered using $\mc{C}_{col}$.  We repeat this
process until all the $m'$ chunks with sector failures are recovered.

\textbf{(3) \emph{Decoding of the $m$ failed chunks:}} After all the $m'$
chunks with sector failures are recovered, the $m$ failed chunks can be
recovered row by row using $\mc{C}_{row}$.

\subsection{Decoding in Practice}
\label{subsec:practice}

In \S\ref{subsec:recovery_proof}, we describe an upstairs decoding method for
the worst case.   In practice, we often have fewer lost symbols than the
worst case defined by $m$ and $\mb{e}$.  To achieve efficient decoding, our
idea is to recover as many lost
symbols as possible via row parity symbols. The reason is that such decoding
is local and involves only the symbols of the same row, while decoding via
global parity symbols involves almost all data symbols within the stripe.  In
our implementation, we first locally recover any lost symbols using row parity
symbols whenever possible. Then, for each chunk that still contains lost
symbols, we count the number of its remaining lost symbols.  Next, we globally
recover the lost symbols with global parity symbols using upstairs decoding
as described in \S\ref{subsec:recovery_proof}, except those in the $m$ chunks
that have the most lost symbols.  These $m$ chunks can be finally recovered
via row parity symbols after all other lost symbols have been recovered.

\section{Extended Encoding: Relocating Global Parity Symbols Inside a Stripe}
\label{sec:transformation}

We thus far assume that there are always $s$ available global parity symbols
that are kept outside a stripe. However, to maintain the regularity of the
code structure and to avoid provisioning extra devices for keeping the global
parity symbols, it is desirable to keep all global parity symbols inside a
stripe.  The idea is that in each stripe, we store the global parity symbols in
some sectors that originally store the data symbols.
A challenge is that such {\em inside global
parity symbols} are also subject to both device and sector failures, so we
must maintain their fault tolerance during encoding.  In this section, we
propose two encoding methods, namely {\em upstairs encoding} and
{\em downstairs encoding}, which support the construction of inside global
parity symbols, while preserving the homomorphic property and hence the fault
tolerance of STAIR codes.  These two encoding
methods produce the same values for parity symbols, but differ in
computational complexities for different configurations. We show how to deduce
parity relations from the two encoding methods, and also show that the two
encoding methods have complementary performance advantages for different
configurations.


\subsection{Two New Encoding Methods}

\subsubsection{Upstairs Encoding}
\label{subsec:upstairs_encoding}

\begin{figure}[t]
\centering
\includegraphics[width=4in]{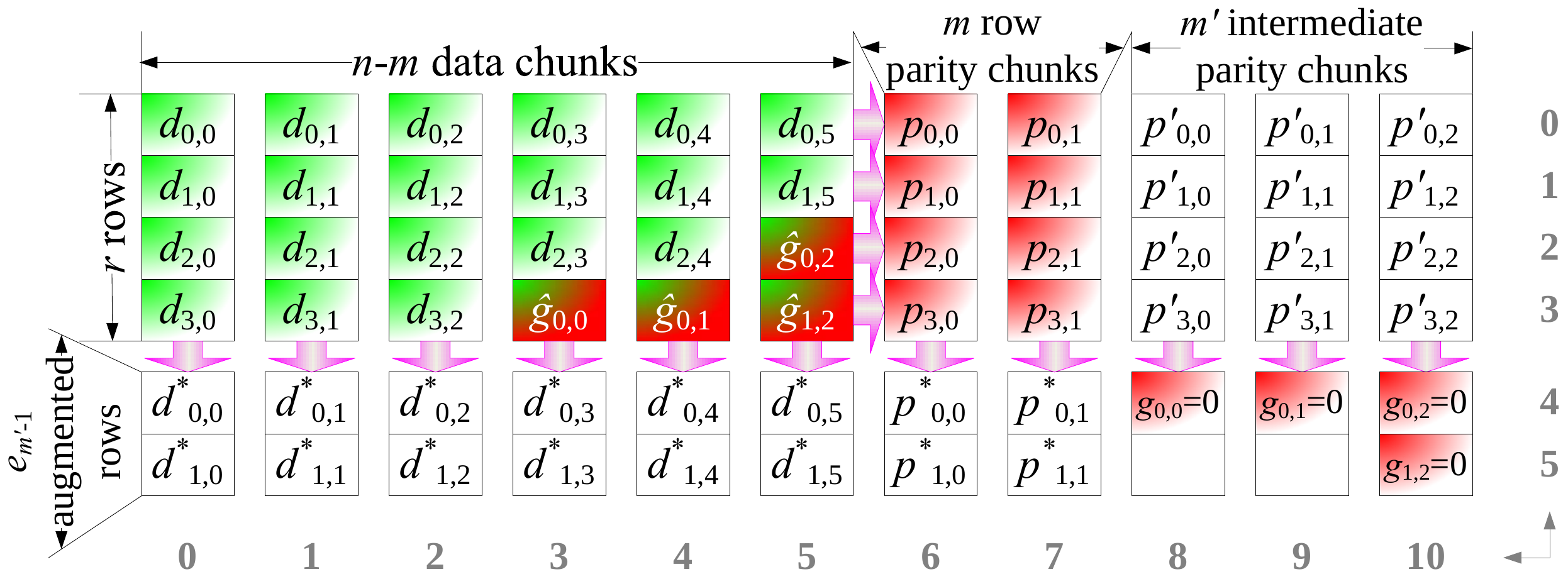}
\caption{Upstairs encoding: we set outside global parity symbols to be zero and
reconstruct the inside global parity symbols using upstairs decoding (see
\S\ref{subsec:recovery_proof}).}
\label{fig:transformation}
\end{figure}

We let $\hat{g}_{h,l}$ ($0\le l\le m'-1$ and $0\le h\le e_l-1$) be an inside
global parity symbol.  Figure~\ref{fig:transformation} illustrates how we
place the inside global parity symbols.  Without loss of generality, we place
them at the bottom of the rightmost data chunks, following the stair layout.
Specifically, we choose the $m'=3$ rightmost data chunks in Columns~3-5
and place $e_0=1$, $e_1=1$, and $e_2=2$ global parity symbols at the bottom of
these data chunks, respectively.  That is, the original data
symbols $d_{3,3}$, $d_{3,4}$, $d_{2,5}$, and $d_{3,5}$ are now replaced by the
inside global parity symbols $\hat{g}_{0,0}$, $\hat{g}_{0,1}$,
$\hat{g}_{0,2}$, and $\hat{g}_{1,2}$, respectively.

\begin{figure}[t]
\centering
\includegraphics[width=4in]{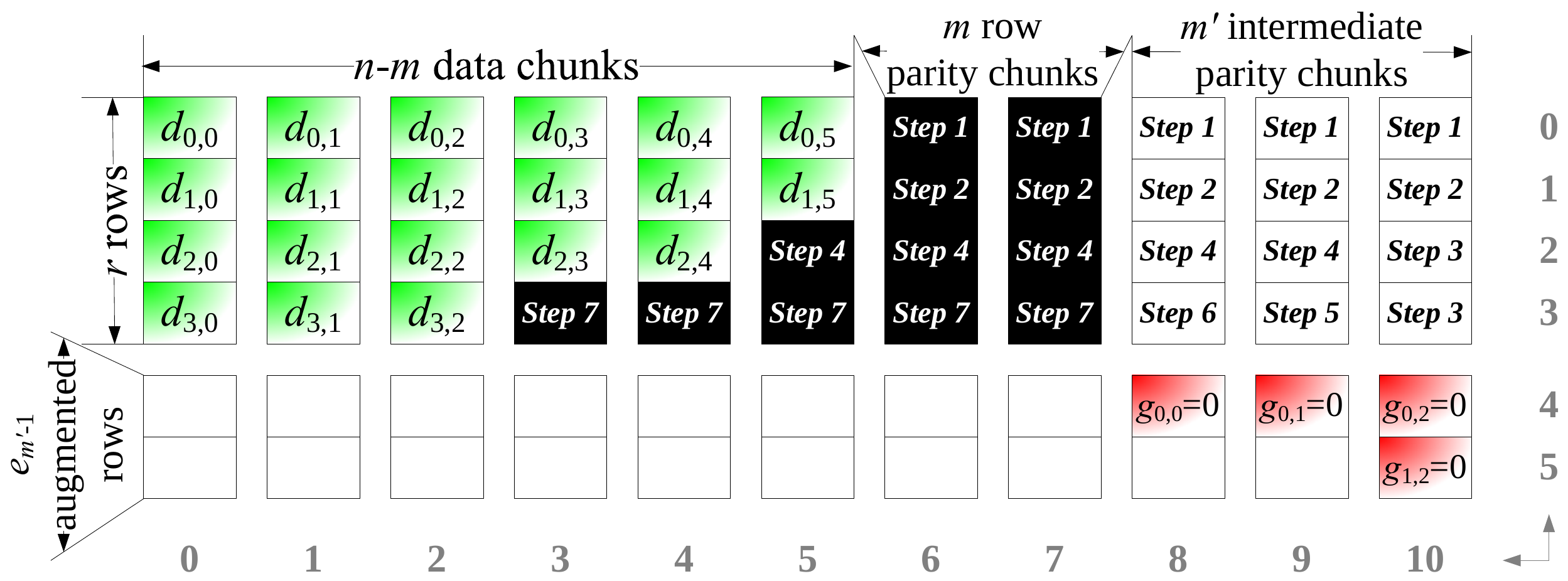}
\caption{Downstairs encoding: we compute the parity symbols from top to bottom
and right to left.}
\label{fig:recover_step_down}
\end{figure}

To obtain the inside global parity symbols, we extend the upstairs decoding
method in \S\ref{subsec:recovery_proof} and propose a recovery-based
encoding approach called {\em upstairs encoding}. We first set all the
outside global parity symbols to be zero (see Figure~\ref{fig:transformation}).
Then we treat all $m=2$ row parity chunks and all $s=4$ inside global parity
symbols as lost chunks and lost sectors, respectively.  Now we ``recover''
all inside global parity symbols, followed by the $m=2$ row parity chunks,
using the upstairs decoding method in \S\ref{subsec:recovery_proof}.  Since
all outside global parity symbols are set to be zero, we need not store
them.  The homomorphic property, and hence the fault tolerance property,
remain the same as discussed in \S\ref{sec:recovery}.  Thus, in failure mode,
we can still use upstairs decoding to reconstruct lost symbols.  We
call this encoding method ``upstairs encoding'' because the parity
symbols are encoded from bottom to top as described in
\S\ref{subsec:recovery_proof}.

\subsubsection{Downstairs Encoding}
\label{subsec:downstairs_encoding}


In addition to upstairs encoding, we present a different encoding method
called {\em downstairs encoding}, in which we generate parity symbols from top
to bottom and right to left.  We
illustrate the idea in Figure~\ref{fig:recover_step_down}, which depicts the
sequence of generating parity symbols.  We still set the outside global parity
symbols to be zero.  First, we encode via $\mc{C}_{row}$ the $n-m=6$ data
symbols in each of the first $r-e_{m'-1} = 2$ rows (i.e., Rows~0 and 1) and
generate $m+m'=5$ parity symbols (including two row parity symbols and three
intermediate parity symbols) (Steps~1-2).  The rightmost
column (i.e., Column~10) now has $r=4$ available symbols, including the two
intermediate parity symbols that are just encoded and two zeroed outside
global parity symbols.  Thus, we can recover $e_{m'-1}=2$ intermediate parity
symbols using $\mc{C}_{col}$ (Step~3).  We can generate $m+m'=5$ parity
symbols (including one inside global parity symbol, two row parity symbols,
and two intermediate parity symbols) for Row~2 using $\mc{C}_{row}$ (Step~4),
followed by $e_{m'-2}=1$ and $e_{m'-3}=1$ intermediate parity symbols in
Columns~9 and 8 using $\mc{C}_{col}$, respectively (Steps~5-6).  Finally, we
obtain the remaining $m+m'=5$ parity symbols (including three global parity
symbols and two row parity symbols) for Row~3 using $\mc{C}_{row}$ (Step~7).
Table~\ref{tab:downstairs} shows the detailed steps of downstairs encoding for
the example in Figure~\ref{fig:recover_step_down}.

\begin{table}[t]
  \centering
  \caption{Downstairs decoding: detailed steps for the example in Figure~\ref{fig:recover_step_down}.}
  \label{tab:downstairs}

\begin{tabular}{|c|c@{\ }c@{\ }c|c|}
  \hline
  \textbf{Step} & \multicolumn{3}{c|}{\textbf{Detailed Description}} & \textbf{Coding Scheme} \\
  \hline
  \hline
  1 & $d_{0,0}, d_{0,1}, d_{0,2}, d_{0,3}, d_{0,4}, d_{0,5}$  & $\Rightarrow$ & $p_{0,0}, p_{0,1}, p'_{0,0}, p'_{0,1}, p'_{0,2}$ & $\mc{C}_{row}$ \\
  \hline
  2 & $d_{1,0}, d_{1,1}, d_{1,2}, d_{1,3}, d_{1,4}, d_{1,5}$  & $\Rightarrow$ &  $p_{1,0}, p_{1,1}, p'_{1,0}, p'_{1,1}, p'_{1,2}$ & $\mc{C}_{row}$ \\
  \hline
  3 & \makecell{$p'_{0,2},p'_{1,2},g_{0,2}=0,g_{1,2}=0$} & $\Rightarrow$ & $p'_{2,2}, p'_{3,2}$ & $\mc{C}_{col}$ \\
  \hline
  4 & \makecell{$d_{2,0}, d_{2,1}, d_{2,2}, d_{2,3}, d_{2,4}, p'_{2,2}$} & $\Rightarrow$ &  $\hat{g}_{0,2}, p_{2,0}, p_{2,1}, p'_{2,0}, p'_{2,1}$ & $\mc{C}_{row}$ \\
  \hline
  5 & \makecell{$p'_{0,1}, p'_{1,1}, p'_{2,1}, g_{0,1}=0$} & $\Rightarrow$ & $p'_{3,1}$ & $\mc{C}_{col}$  \\
  \hline
  6 & \makecell{$p'_{0,0}, p'_{1,0}, p'_{2,0}, g_{0,0}=0$} & $\Rightarrow$ & $p'_{3,0}$ & $\mc{C}_{col}$  \\
  \hline
  7 & \makecell{$d_{3,0}, d_{3,1}, d_{3,2}, p'_{3,0}, p'_{3,1}, p'_{3,2}$} & $\Rightarrow$ &  $\hat{g}_{0,0}, \hat{g}_{0,1}, \hat{g}_{1,2}, p_{3,0}, p_{3,1}$ & $\mc{C}_{row}$ \\
  \hline
\end{tabular}
\end{table}

In general, we start with encoding via $\mc{C}_{row}$ the rows from top to
bottom.  In each row, we generate $m+m'$ symbols. When no more rows can
be encoded because of insufficient available symbols, we encode via
$\mc{C}_{col}$ the columns from right to left to obtain new intermediate
parity symbols (initially, we obtain $e_{m'-1}$ symbols, followed by
$e_{m'-2}$ symbols, and so on).  We alternately encode rows and columns until
all parity symbols are formed.   We can generalize the steps as in
\S\ref{subsec:general}, but we omit the details in the interest of space.

It is important to note that the downstairs encoding method cannot be
generalized for decoding lost symbols.
For example, referring to our exemplary configuration, we consider a
worst-case recoverable failure scenario in which both row parity chunks are
entirely failed, and the data symbols $d_{0,3}$, $d_{1,4}$, $d_{2,2}$, and
$d_{3,2}$ are lost.  In this case, we cannot recover the lost symbols in the
top row first, but instead we must resort to upstairs decoding as described in
\S\ref{subsec:recovery_proof}.  Upstairs decoding works because we limit the
maximum number of chunks with lost symbols (i.e., at most $m+m'$).  This
enables us to first recover the leftmost virtual parity symbols of the
augmented rows first and gradually reconstruct lost symbols.  On the other
hand, we do not limit the number of rows with lost symbols in our
configuration, so the downstairs method cannot be used for general decoding.

\subsubsection{Discussion}

Note that both upstairs and downstairs encoding methods always generate the
{\em same} values for all parity symbols, since both of them preserve the
homomorphic property, fix the outside global parity symbols to be zero, and use
the same schemes $\mc{C}_{row}$ and $\mc{C}_{col}$ for encoding.

Also, both of them reuse parity symbols in the intermediate steps
to generate additional parity symbols in subsequent steps.  On the other hand,
they differ in encoding complexity, due to the different ways of reusing
the parity symbols.  We analyze this in \S\ref{subsec:analysis}.

\subsection{Uneven Parity Relations}
\label{subsec:uneven}

Before relocating the global parity symbols inside a stripe,
each data symbol contributes to $m$ row parity symbols and all $s$ outside
global parity symbols.  However, after relocation, the parity relations become
uneven. That is, some row parity symbols are also contributed by the data
symbols in other rows, while some inside global parity symbols are contributed
by only a subset of data symbols in the stripe.  Here, we discuss the uneven
parity relations of STAIR codes so as to better understand the encoding and
update performance of STAIR codes in subsequent analysis.



\begin{figure}[t]
\centering
\includegraphics[width=2.2in]{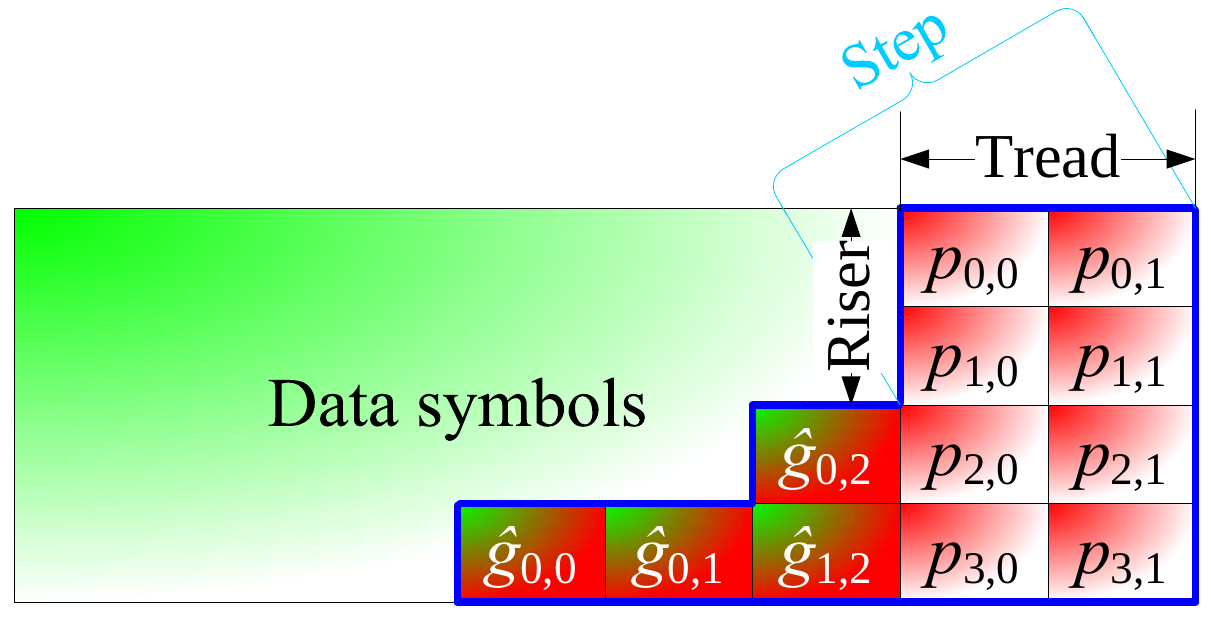}
\caption{A stair step with a tread and a riser.}
\label{fig:stair}
\end{figure}

To analyze how exactly each parity symbol is generated, we revisit both
upstairs and downstairs encoding methods.  Recall that the row parity symbols
and the inside global parity symbols are arranged in the form of stair steps,
each of which is composed of a {\em tread} (i.e., the horizontal portion of a
step) and a {\em riser} (i.e., the vertical portion of a step), as shown in
Figure~\ref{fig:stair}.  If upstairs encoding is used, then from
Figure~\ref{fig:recover_step}, the encoding of each parity symbol does not
involve any data symbol on its right.  Also, among the columns spanned by the
same tread, the encoding of parity symbols in each column does not
involve any data symbol in other columns.  We can make similar arguments for
downstairs encoding.  If downstairs encoding is used, then from
Figure~\ref{fig:recover_step_down}, the encoding of each parity symbol does
not involve any data symbol below it.  Also, among the rows spanned by the
same riser, the encoding of parity symbols in each row does not involve
any data symbol in other rows.

\begin{figure}[t]
\centering
\includegraphics[width=5.5in]{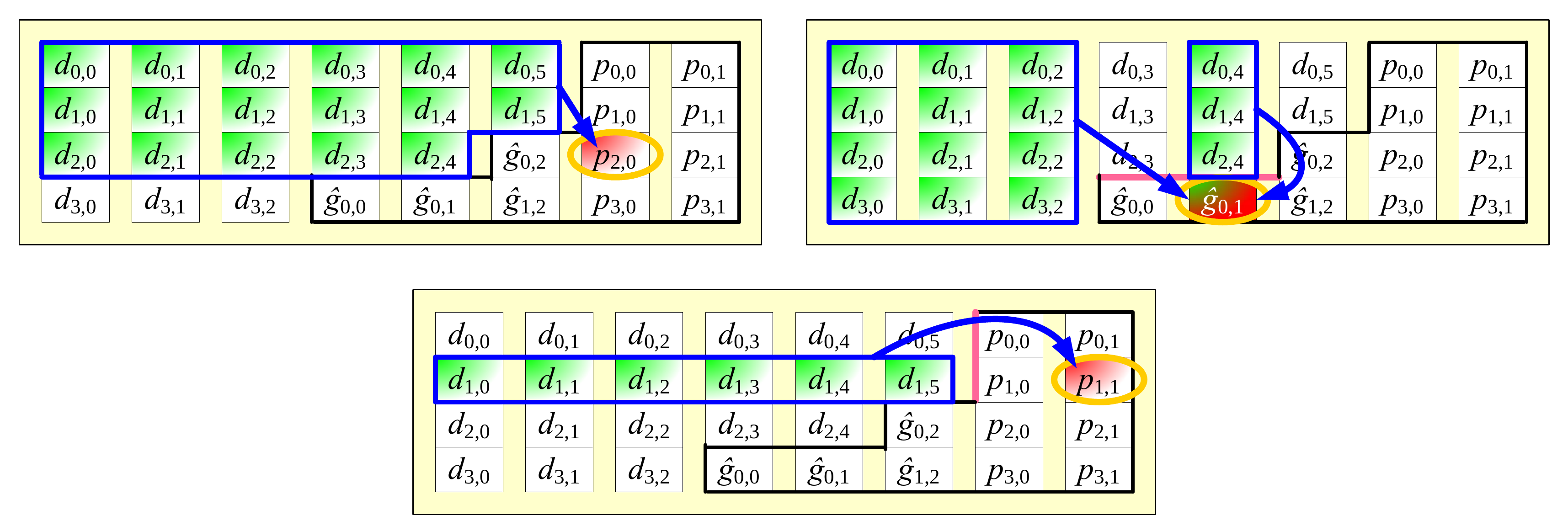}
\caption{The data symbols that contribute to parity symbols $p_{2,0}$,
$\hat{g}_{0,1}$, and $p_{1,1}$, respectively.}
\label{fig:uneven_parity_relations}
\end{figure}

As both upstairs and downstairs encoding methods generate the same values of
parity symbols, we can combine the above arguments into the following
property of how each parity symbol is related to data symbols.
\begin{property}\label{obs:parity_relation}
\rm
{\bf (Parity relations in STAIR codes):} In a STAIR code stripe, a
(row or inside global) parity symbol in Row~$i_0$ and Column~$j_0$ (where
$0\leq i_0 \leq r-1$ and $n-m-m' \leq j_0 \leq n-1$) depends only on the
data symbols $d_{i,j}$'s where $i\leq i_0$ and $j\leq j_0$.  Moreover,
each parity symbol is unrelated to any data symbol in any
other column (row) spanned by the same tread (riser).
\end{property}

Figure~\ref{fig:uneven_parity_relations} illustrates the above property.
For example, $p_{2,0}$ depends only on the data symbols
$d_{i,j}$'s in Rows~0-2 and Columns~0-5.  Note that
$\hat{g}_{0,1}$ in Column~4 is unrelated to any data symbol in Column~3,
which is spanned by the same tread as Column~4. Similarly, $p_{1,1}$
in Row~1 is unrelated to any data symbol in Row~0, which is spanned by the
same riser as Row~1.

\subsection{Encoding Complexity Analysis}
\label{subsec:analysis}

\begin{figure}[t]
\centering
\includegraphics[width=5.2in]{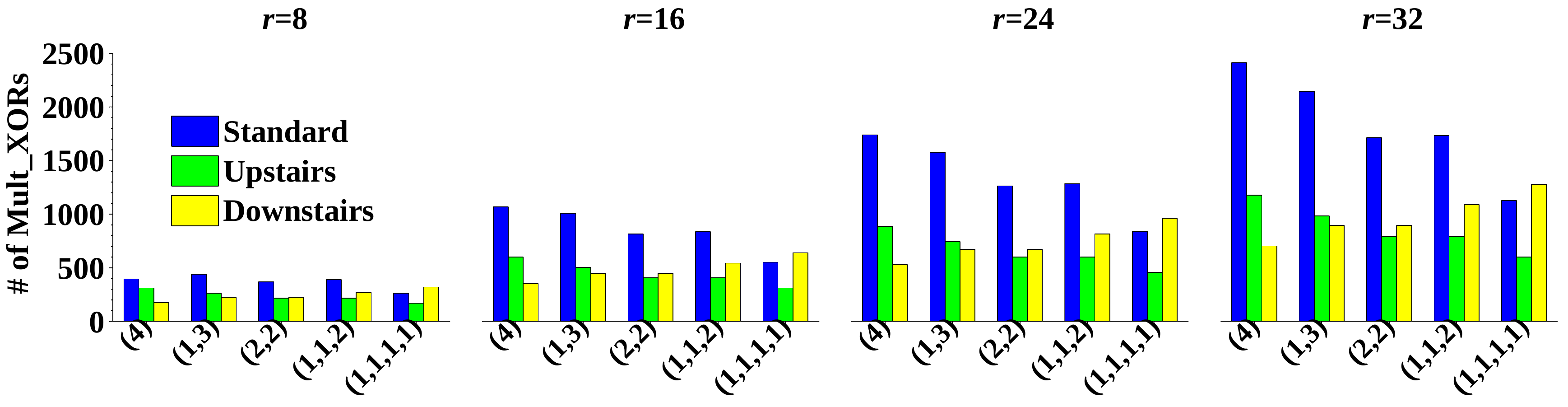}
\caption{Numbers of \texttt{Mult\_XOR}s (per stripe) of the three
encoding methods for STAIR codes versus different $\mb{e}$'s when $n=8$,
 $m=2$, and $s=4$.}
\label{fig:different_encoding_methods}
\end{figure}

We have proposed two encoding methods for STAIR codes: upstairs encoding and
downstairs encoding.  Both of them alternately encode rows and columns to
obtain the parity symbols.  We can also obtain parity symbols using the
standard encoding approach, in which each parity symbol is computed directly
from a linear combination of data symbols as in classical Reed-Solomon codes.
We now analyze the computational complexities of these three methods for
different configuration parameters of STAIR codes.

STAIR codes perform encoding over a Galois Field, in which linear arithmetic
can be decomposed into the basic operations \texttt{Mult\_XOR}s
\cite{Plank13d}.  We define
$\mbox{\texttt{Mult\_XOR}}(\mb{R}_1, \mb{R}_2, a)$ as an operation that
first multiplies a region $\mb{R}_1$ of bytes by a $w$-bit constant $a$
in Galois Field $GF(2^w)$, and then applies XOR-summing to the product and the
target region $\mb{R}_2$ of the same size.
For example, $\mb{Y}=a_0\cdot\mb{X}_0+a_1\cdot\mb{X}_1$ can be
decomposed into two \texttt{Mult\_XOR}s (assuming $\mb{Y}$ is initialized as
zero): $\mbox{\texttt{Mult\_XOR}}(\mb{X}_0, \mb{Y}, a_0)$ and
$\mbox{\texttt{Mult\_XOR}}(\mb{X}_1, \mb{Y}, a_1)$.
Clearly, fewer \texttt{Mult\_XOR}s imply a lower computational complexity. To
evaluate the computational complexity of an encoding method, we count its
number of \texttt{Mult\_XOR}s (per stripe).

For upstairs encoding, we generate $m\cdot r$ row parity symbols and $s$
virtual parity symbols along the row direction, as well as $s$ inside global
parity symbols and $(n-m)\cdot e_{m'-1}-s$ virtual parity symbols along the
column direction.  Its number of \texttt{Mult\_XOR}s (denoted by
$\mc{X}_{up}$) is:
\begin{equation}
\mc{X}_{up} = \overset{\footnotesize \mbox{row direction}}{\overbrace{(n-m) \times (m \cdot r + s)}} + \overset{\footnotesize \mbox{column direction}}{\overbrace{r \times \left[(n-m)\cdot e_{m'-1}\right]}}.
\label{eqn:upstairs}
\end{equation}

For downstairs encoding, we generate $m\cdot r$ row parity symbols,
$s$ inside global parity symbols, and $m'\cdot r-s$ intermediate parity
symbols along the row direction, as well as $s$ intermediate parity symbols
along the column direction. Its number of $\texttt{Mult\_XOR}$s (denoted by
$\mc{X}_{down}$) is:
\begin{equation}
\mc{X}_{down} = \overset{\footnotesize \mbox{row direction}}{\overbrace{(n-m) \times \left[(m+m') \cdot r\right]}} + \overset{\footnotesize \mbox{column direction}}{\overbrace{r \times s}}.
\label{eqn:downstairs}
\end{equation}


For standard encoding, we compute the number of $\texttt{Mult\_XOR}$s by
summing the number of data symbols that contribute to each parity
symbol, based on the property of uneven parity relations discussed in
\S\ref{subsec:uneven}.

%


We show via a case study how the three encoding methods differ in the number
of \texttt{Mult\_XOR}s.
Figure~\ref{fig:different_encoding_methods} depicts the numbers
of \texttt{Mult\_XOR}s of the three encoding methods for different $\mb{e}$'s
in the case where $n=8$, $m=2$, and $s=4$.  Upstairs encoding and downstairs
encoding incur significantly fewer \texttt{Mult\_XOR}s than standard
encoding most of the time. The main reason is that both upstairs encoding and
downstairs encoding often reuse the computed parity symbols in subsequent
encoding steps.
We also observe that for a given $s$, the number of \texttt{Mult\_XOR}s of
upstairs encoding increases with $e_{m'-1}$ (see
Equation~(\ref{eqn:upstairs})), while that of downstairs encoding increases
with $m'$ (see Equation~(\ref{eqn:downstairs})). Since larger $m'$ often
implies smaller $e_{m'-1}$, the value of $m'$ often determines which of the
two encoding methods is more efficient: when $m'$ is small, downstairs
encoding wins; when $m'$ is large, upstairs encoding wins.


In our encoding implementation of STAIR codes, for given configuration
parameters, we always pre-compute the number of \texttt{Mult\_XOR}s for each
of the encoding methods, and then choose the one with the fewest
\texttt{Mult\_XOR}s.

\section{Storage and Performance Evaluation}
\label{sec:evaluation}

\begin{figure}[t]
\centering
\includegraphics[width=5.2in]{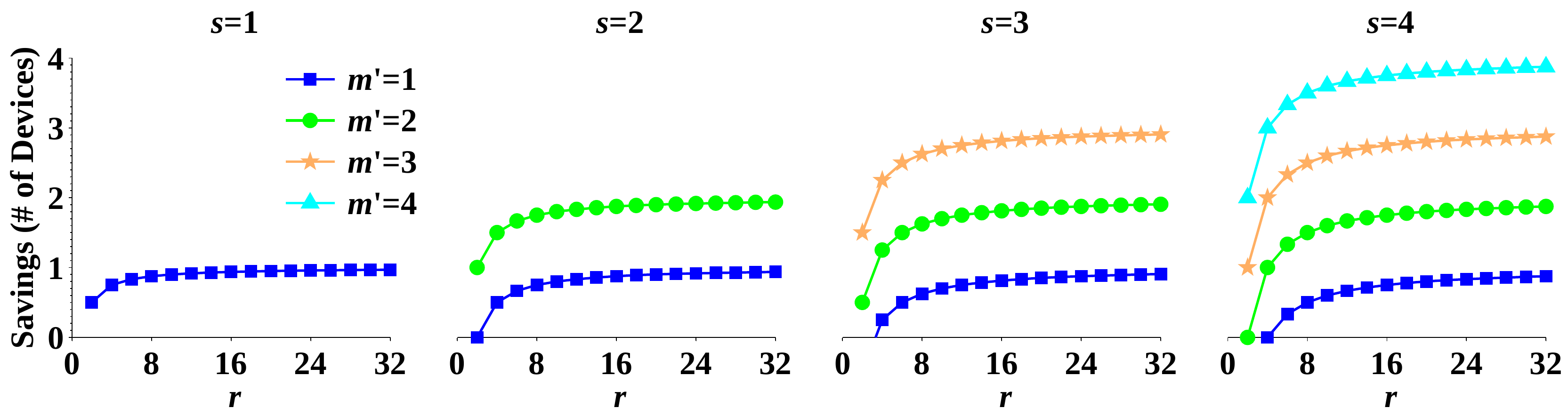}
\caption{Space saving of STAIR codes over traditional erasure codes in terms
of $s$, $m'$, and $r$.}
\label{fig:space_saving}
\end{figure}

We evaluate STAIR codes and compare them with other related erasure codes in
different practical aspects, including storage space saving, encoding/decoding
speed, and update penalty.

\subsection{Storage Space Saving}


The main motivation for STAIR codes is to tolerate simultaneous device and
sector failures with significantly lower storage space overhead than
traditional erasure codes (e.g., Reed-Solomon codes) that provide only
device-level fault tolerance.  Given a failure scenario defined by $m$ and
$\mb{e}$, traditional erasure codes need $m+m'$ chunks per stripe for parity,
while STAIR codes need only $m$ chunks and $s$ symbols (where $m'\leq s$).
Thus, STAIR codes save $r\times m'-s$ symbols per stripe, or equivalently,
$m'-\frac{s}{r}$ devices per system.  In short, the saving of STAIR codes
depends on only three parameters $s$, $m'$, and $r$ (where $s$ and $m'$ are
determined by $\mb{e}$).


Figure~\ref{fig:space_saving} plots the number of devices saved by STAIR codes
for $s \leq 4$, $m'\le s$, and $r \leq 32$.  As $r$ increases, the number of
devices saved is close to $m'$.  The saving reaches the highest when $m'=s$.

We point out that the recently proposed SD codes \cite{Plank13b,Plank13c} are
also motivated for reducing the storage space over traditional erasure codes.
Unlike STAIR codes, SD codes always achieve a saving of $s-\frac{s}{r}$
devices, which is the maximum saving of STAIR codes.  While STAIR codes
apparently cannot outperform SD codes in space saving, it is important to note
that the currently known constructions of SD codes are limited to $s\leq 3$
only \cite{Plank13b,Plank13c,Blaum13c}, implying that SD codes can save no more than
three devices.  On the other hand, STAIR codes do not have such limitations.
As shown in Figure~\ref{fig:space_saving}, STAIR codes can save more than
three devices for larger $s$.

\subsection{Encoding/Decoding Speed}

We evaluate the encoding/decoding speed of STAIR codes.  Our implementation of
STAIR codes is written in C.  We leverage the GF-Complete open source library
\cite{Plank13d} to accelerate Galois Field arithmetic using Intel SIMD
instructions.  Our experiments compare STAIR codes with the state-of-the-art
SD codes \cite{Plank13b,Plank13c}. At the time of this writing, the
open-source implementation of SD codes encodes stripes in a decoding manner
without any parity reuse. For fair comparisons, we extend the SD code
implementation to support the standard encoding method mentioned in
\S\ref{subsec:analysis}.
We run our performance tests on a machine equipped with an Intel Core i5-3570
CPU at 3.40GHz with SSE4.2 support. The CPU has a 256KB L2-cache and a 6MB
L3-cache.

\subsubsection{Encoding}
\label{subsec:encoding_perf}


We compare the encoding performance of STAIR codes and SD codes for different
values of $n$, $r$, $m$, and $s$.  For SD codes, we only consider the range of
configuration parameters where $s\leq 3$, since no code construction is
available outside this range \cite{Plank13b,Plank13c,Blaum13c}.
In addition, the SD code constructions for $s=3$ are only available in the
range $n\leq 24$, $r\leq 24$, and $m\le 3$ \cite{Plank13b,Plank13c}.
For STAIR codes, a single value of $s$ can imply different configurations of
$\mb{e}$ (e.g., see Figure~\ref{fig:different_encoding_methods} in
\S\ref{subsec:analysis}), each of which has different encoding
performance.  Here, we take a conservative approach to analyze the
worst-case performance of STAIR codes, that is, we test all possible
configurations of $\mb{e}$ for a given $s$ and pick the one with the
lowest encoding speed.

Note that the encoding performance of both STAIR codes and SD codes heavily
depends on the word size $w$ of the adopted Galois Field $GF(2^w)$, where $w$
is often set to be a power of 2.  A smaller $w$ often means a higher encoding
speed \cite{Plank13d}.  STAIR codes work as long as $n+m'\leq 2^w$ and
$r+e_{m'-1}\leq 2^w$.  Thus, we choose $w = 8$ since it suffices for all of
our tests.  However, SD codes may choose among $w = 8$, $w = 16$, and
$w = 32$, depending on configuration parameters.  We choose the smallest $w$
that is feasible for the SD code construction.

We consider the metric {\em encoding speed}, defined as the amount of data
encoded per second.  We construct a stripe of size roughly 32MB in memory 
\cite{Plank13b,Plank13c}.  We put random bytes in the stripe, and divide
the stripe into $r\times n$ sectors, each mapped to a symbol.  We obtain the
averaged results over 10 runs.

	


\begin{figure}[t]
  \centering
  \subfigure[Varying $n$ when $r=16$]{
    \label{fig:encoding_speed_n}
    \includegraphics[width=5in]{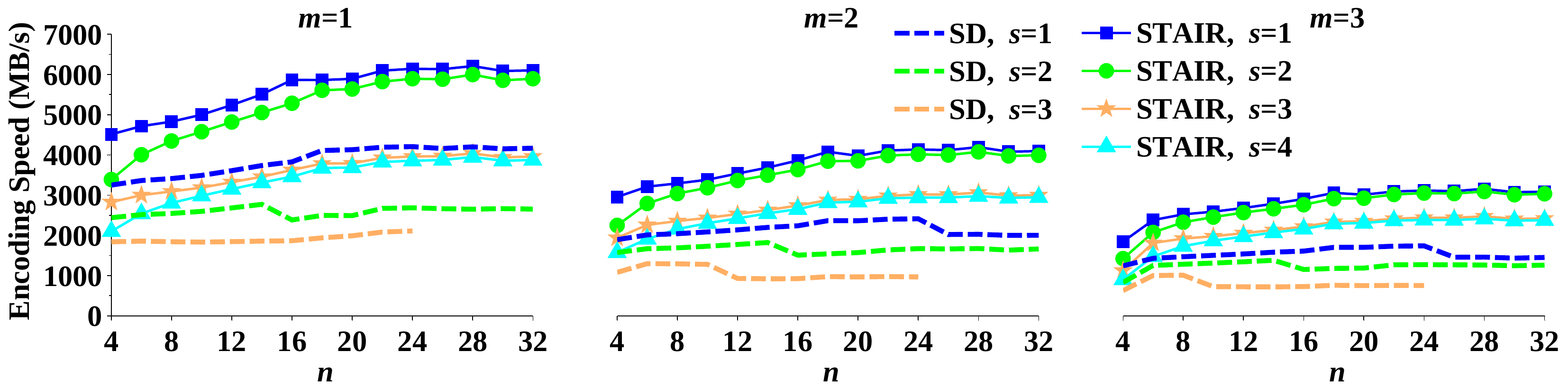}
  }
  \hspace{1in}
  \subfigure[Varying $r$ when $n=16$]{
    \label{fig:encoding_speed_r}
    \includegraphics[width=5in]{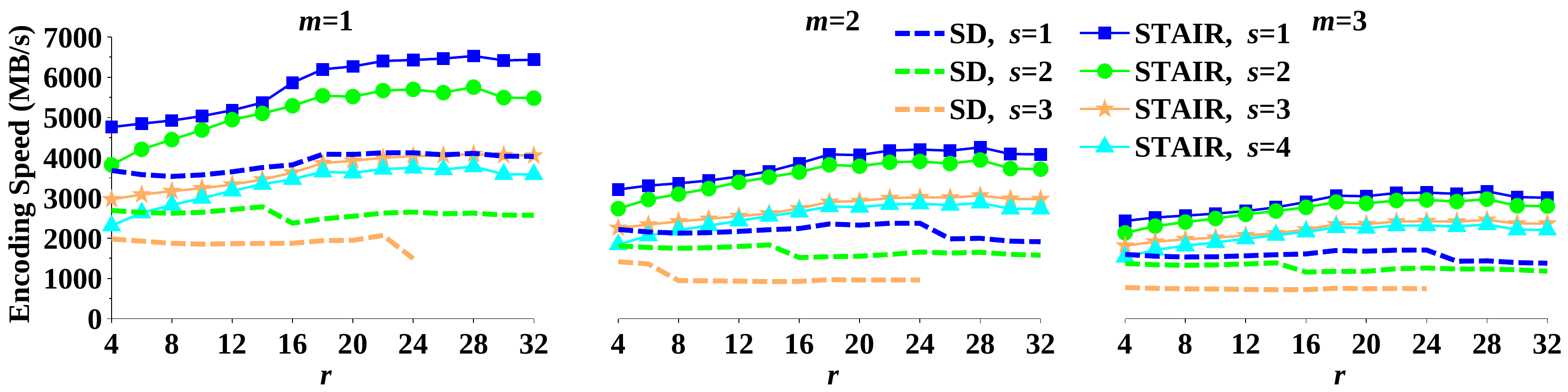}
  }
  \caption{Encoding speed of STAIR codes and SD codes for different
	  combinations of $n$, $r$, $m$, and $s$.}
  \label{fig:encoding_speed}
\end{figure}

\begin{figure}[t]
\centering
\includegraphics[width=5in]{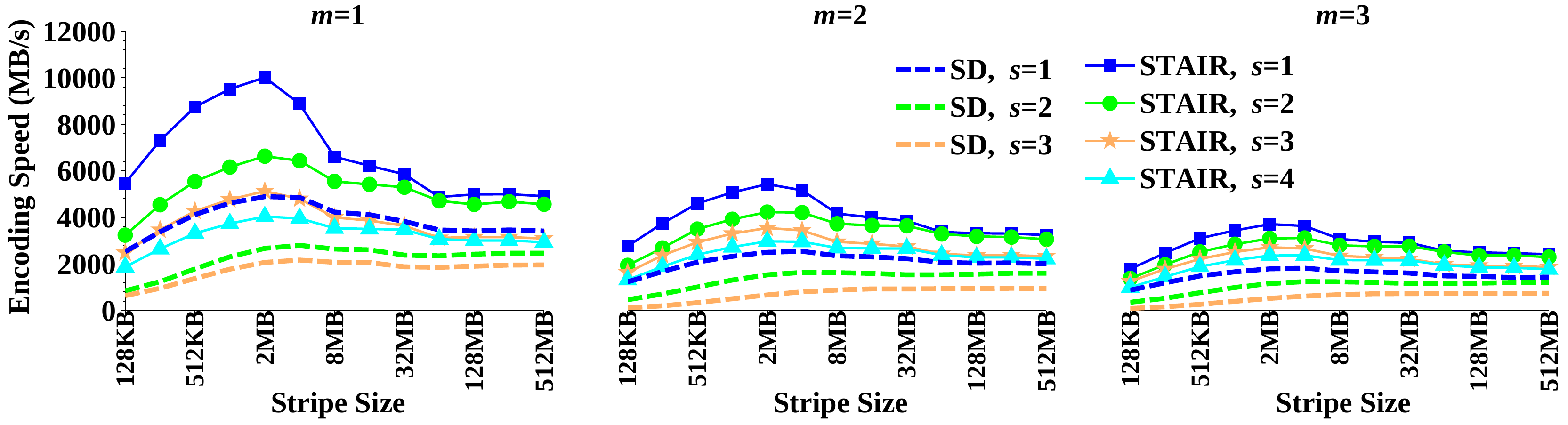}
\caption{Encoding speed of STAIR codes and SD codes for different stripe sizes
when $n=16$ and $r=16$.}
\label{fig:encoding_stripe}
\end{figure}


Figures~\ref{fig:encoding_speed}(a) and \ref{fig:encoding_speed}(b) present
the encoding speed results for different values of $n$ when $r = 16$ and for
different values of $r$ when $n = 16$, respectively.  In most cases, the
encoding speed of STAIR codes is over 1000MB/s, which is significantly higher
than the disk write speed in practice (note that although disk writes can be
parallelized in disk arrays, the encoding operations can also be parallelized
with modern multi-core CPUs). The speed increases with both $n$ and $r$. The
intuitive reason is that the proportion of parity symbols decreases with $n$
and $r$.
Compared to SD codes, STAIR codes improve the encoding speed by $106.03\%$
on average (in the range from $29.30\%$ to $225.14\%$).
The reason is that STAIR codes reuse encoded parity information in subsequent encoding steps by
upstairs/downstairs encoding (see \S\ref{subsec:analysis}), while such an
encoding property is not exploited in SD codes.


We also evaluate the impact of stripe size on the encoding speed of STAIR
codes and SD codes for given $n$ and $r$.  We fix $n=16$ and $r=16$, and vary
the stripe size from 128KB to 512MB.
Note that a stripe of size 128KB implies
a symbol of size 512~bytes, the standard sector size in practical disk drives.
Figure~\ref{fig:encoding_stripe} presents the encoding speed results. As the
stripe size increases, the encoding speed of both STAIR codes and SD codes
first increases and then drops, due to the mixed effects of SIMD instructions
adopted in GF-Complete \cite{Plank13d} and CPU cache.  Nevertheless, the
encoding speed advantage of STAIR codes over SD codes remains unchanged.

\subsubsection{Decoding}
\label{subsec:decoding_perf}

\begin{figure}[t]
  \centering
  \subfigure[Varying $n$ when $r=16$]{
    \label{fig:decoding_speed_r}
    \includegraphics[width=5in]{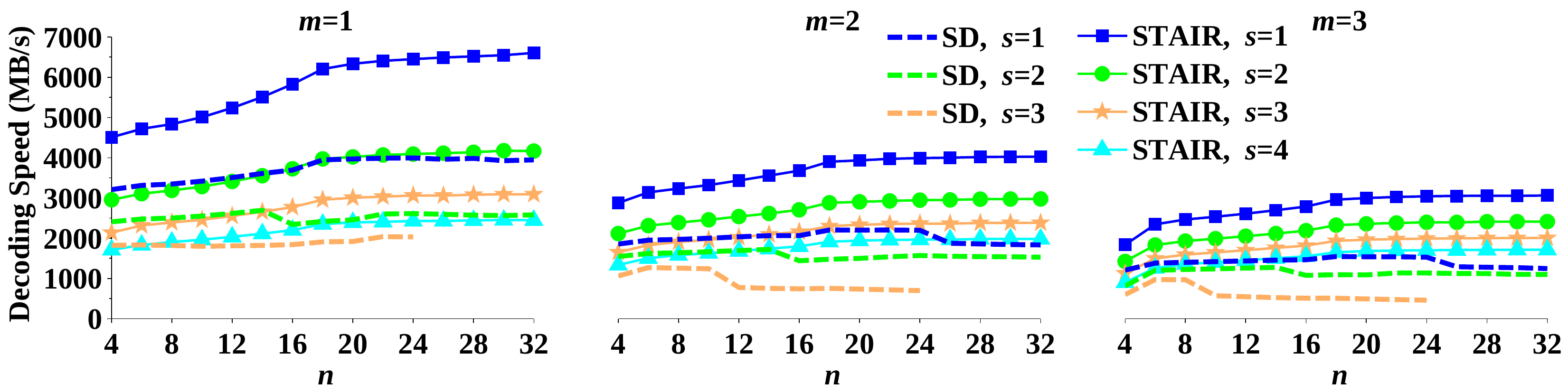}
  }
  \hspace{1in}
  \subfigure[Varying $r$ when $n=16$]{
    \label{fig:decoding_speed_n}
    \includegraphics[width=5in]{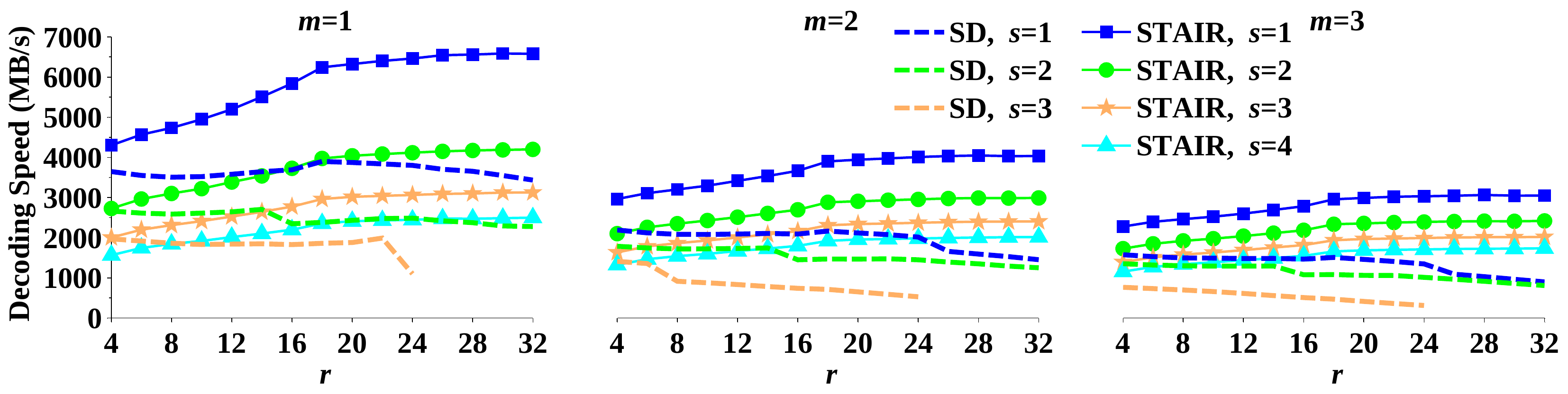}
  }
  \caption{Decoding speed of STAIR codes and SD codes for different
	  combinations of $n$, $r$, $m$, and $s$.}
  \label{fig:decoding_speed}
\end{figure}

We measure the decoding performance of STAIR codes and SD codes in recovering
lost symbols. Since the decoding time increases with the number of lost
symbols to be recovered, we consider a particular worst case in which the $m$
leftmost chunks and $s$ additional symbols in the following $m'$ chunks
defined by $\mb{e}$ are all lost.  The evaluation setup is similar to that in
\S\ref{subsec:encoding_perf}, and in particular, the stripe size is fixed at
32MB.

Figures~\ref{fig:decoding_speed_r} and \ref{fig:decoding_speed_n} present the
decoding speed results for different $n$ when $r=16$ and for
different $r$ when $n=16$, respectively.  The results of both
figures can be viewed in comparison to those of
Figures~\ref{fig:encoding_speed_n} and \ref{fig:encoding_speed_r},
respectively.  Similar to encoding, the decoding speed of STAIR codes is
over 1000MB/s in most cases
and increases with both $n$ and $r$. Compared to SD codes, STAIR codes
improve the decoding speed by $102.99\%$ on average (in the range from $1.70\%$
to $537.87 \%$).

In practice, we often have fewer lost symbols than the worst case (see
\S\ref{subsec:practice}).  One common case is that there are only failed
chunks due to device failures (i.e., $s=0$), so the decoding of both STAIR and
SD codes is identical to that of Reed-Solomon codes.  In this case, the
decoding speed of STAIR/SD codes can be significantly higher than that of
$s=1$ for STAIR codes in Figure~\ref{fig:decoding_speed}. For example, when
$n=16$ and $r=16$, the decoding speed increases by 79.39\%, 29.39\%, and
11.98\% for $m=1$, $2$, and $3$, respectively.

\subsection{Update Penalty}
\label{subsec:update_penalty}


\begin{figure}[t]
\centering
\includegraphics[width=5in]{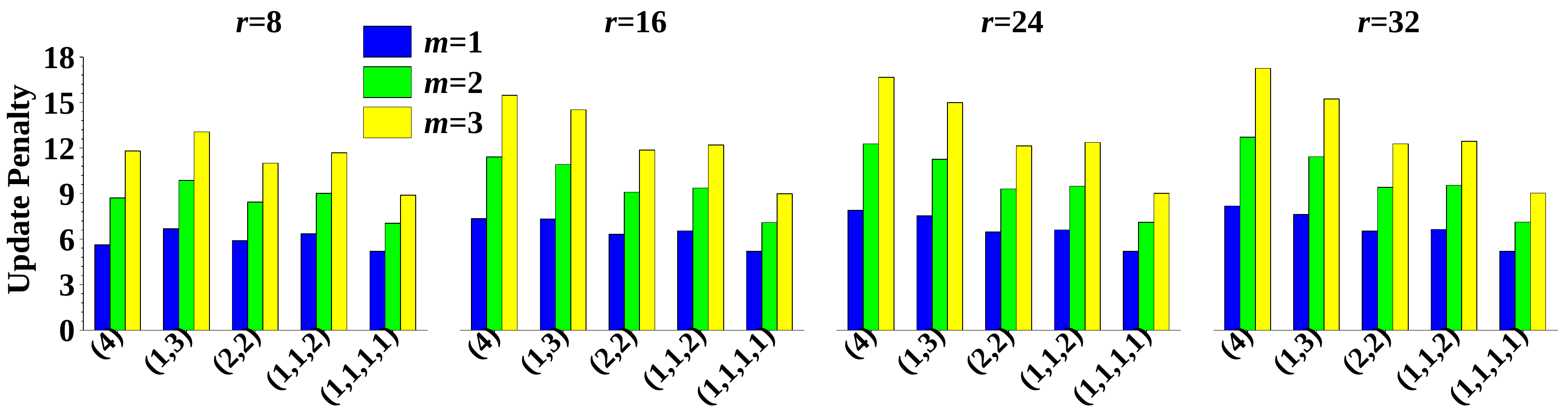}
\caption{Update penalty of STAIR codes for different $\mb{e}$'s when $n=16$
and $s=4$.}
\label{fig:update_error_vector}
\end{figure}

\begin{figure}[t]
\centering
\includegraphics[width=5in]{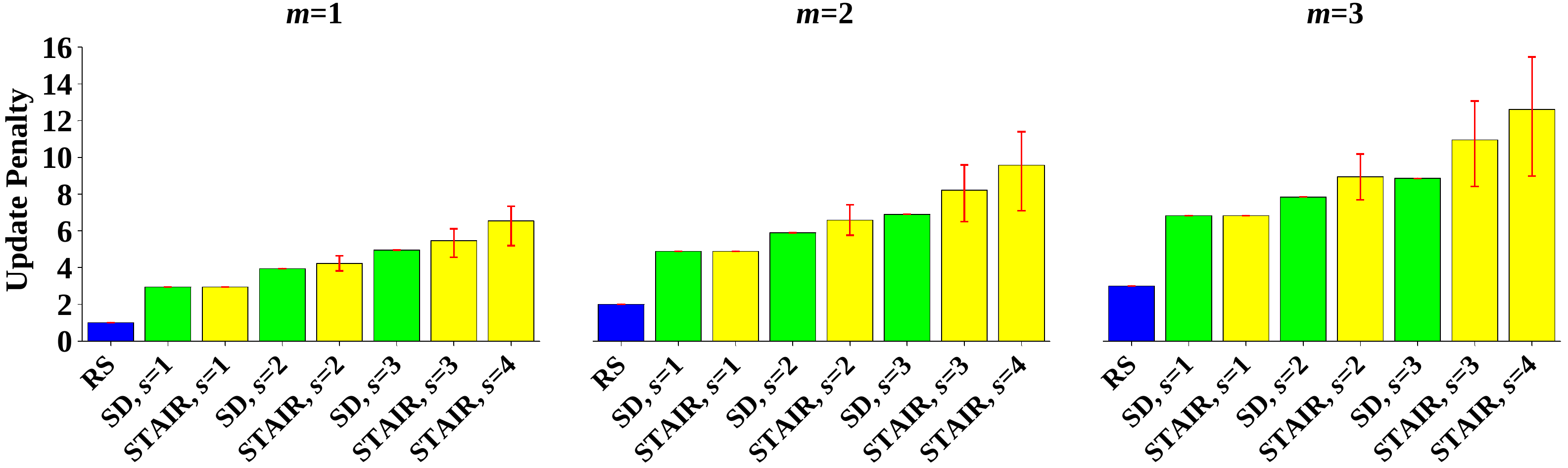}
\caption{Update penalty of STAIR codes, SD codes, and Reed-Solomon (RS) codes
when $n=16$ and $r=16$.  For STAIR codes, we plot the error bars for the
maximum and minimum update penalty values among all possible configurations of
$\mb{e}$.}
\label{fig:update_penalty}
\end{figure}

We evaluate the update cost of STAIR codes when data symbols are
updated.
For each data symbol in a stripe being updated, we count the number of parity
symbols being affected (see \S\ref{subsec:uneven}). Here, we define the
{\em update penalty} as the average number of parity symbols that need to be
updated when a data symbol is updated.



Clearly, the update penalty of STAIR codes increases with $m$. We are more
interested in how $\mb{e}$ influences the update penalty of STAIR codes.
Figure~\ref{fig:update_error_vector} presents the update penalty results for
different $\mb{e}$'s when $n=16$ and $s=4$. For different $\mb{e}$'s with the
same $s$, the update penalty of STAIR codes often increases with $e_{m'-1}$.
Intuitively, a larger $e_{m'-1}$ implies that more rows of row parity symbols
are encoded from inside global parity symbols, which are further encoded from
almost all data symbols (see \S\ref{subsec:uneven}).

We compare STAIR codes with SD codes \cite{Plank13b,Plank13c}.
For STAIR codes with a given $s$, we test all possible
configurations of $\mb{e}$ and find the average, minimum, and maximum update
penalty.  For SD codes, we only consider $s$ between 1 and 3.  We also include
the update penalty results of Reed-Solomon codes for reference.
Figure~\ref{fig:update_penalty} presents the update penalty results when
$n=16$ and $r=16$ (while similar observations are made for other $n$ and $r$).
For a given $s$, the range of update penalty of STAIR codes covers that of SD
codes, although the average is sometimes higher than that of SD codes (same
for $s=1$, by 7.30\% to 14.02\% for $s=2$, and by 10.47\% to 23.72\% for
$s=3$).
Both STAIR codes and SD codes have higher update penalty than Reed-Solomon
codes due to more parity symbols in a stripe, and hence are suitable for
storage systems with rare updates (e.g., backup or write-once-read-many
(WORM) systems) or systems dominated by full-stripe writes
\cite{Plank13b,Plank13c}.

\section{Reliability Analysis}
\label{sec:reliability}

In the previous section, we examine the storage and performance properties of
STAIR codes.  We now characterize the reliability of STAIR codes using
analytical models.  We also show that STAIR codes effectively tolerate sector
failure bursts \cite{Bairavasundaram07,Schroeder10} by supporting a wide range
of configurations of the sector failure coverage defined by $\mb{e}$.   We
extend the reliability analysis by Dholakia {\em et al.} \cite{Dholakia08}
specifically for STAIR codes, whose fault tolerance is defined by the specific
configuration of $\mb{e}$.  Table~\ref{tab:notation-reliability} summarizes
the major notation for our reliability analysis.

\begin{table}[t]
  \centering
  \caption{Major notation used for reliability analysis.}
  \label{tab:notation-reliability}
  \begin{small}
  \begin{tabular}{lp{4.3in}}
    \toprule
    \textbf{Notation} & \hfil \textbf{Description} \hfil \\
    \midrule
    $U$ & Total amount (in bytes) of user data stored in a storage system \\
    $C$ & Device capacity (in bytes) \\
    $S$ & Sector size (in bytes)  \\
    $E$ & Storage efficiency of an erasure code \\
    $N_{arr}$ & Number of storage arrays in a storage system \\
    $MTTDL_{sys}$ & MTTDL of a storage system \\
    ${MTTDL}_{arr}$ & MTTDL of a single storage array \\
    $1/\lambda$ & Mean time to device failure \\
    $1/\mu$ & Mean time to rebuild in critical mode\\
	$P_{arr}$ & Probability that a storage array in critical mode
	encounters unrecoverable sector failures in non-failed devices\\
    $P_{str}$ & Probability that a stripe in critical mode
	encounters unrecoverable sector failures in non-failed chunks \\
	$P_{chk(i)}$ & Probability that a chunk encounters $i$ sector failures
	(where $0 \leq i \leq r$) \\
    $P_{bit}$ & Probability of an unrecoverable bit error\\
    $P_{sec}$ & Probability of a sector failure\\
	$B$ & Average length (in number of sectors) of a sector failure burst\\
	$b_i$ & Fraction of sector failure bursts of length $i$ (where $i \geq 1$)\\
    $\alpha$ &  Tail index of a Pareto distribution that best fits the distribution of length $\geq 2$ for sector failure bursts \\
    \bottomrule
  \end{tabular}
  \end{small}
\end{table}

\subsection{Analytical Models}

In this subsection, we develop analytical models for the reliability analysis.

\subsubsection{MTTDL Model}

We first model the overall reliability of a storage system.  We use the
standard reliability metric called \emph{mean time to data loss (MTTDL)},
although other advanced metrics have been proposed in the literature
\cite{Greenan10}.

Recall from \S\ref{sec:problem} that we encode a storage array using a STAIR
code with configuration parameters $n$, $r$, $m$, and $\mb{e}$ (and hence
$s$).  Consider a storage system with $N_{arr}$ storage arrays, each with
$n$ devices of capacity $C$.  To store a given amount $U$ of user data,
$N_{arr}$ should be set to be:
\begin{equation}\label{equ:n-array}
N_{arr} = \left\lceil \frac{\left. U \middle/ E \right.}{C \cdot n} \right\rceil,
\end{equation}
where $E$ denotes the storage efficiency of an erasure code (i.e., the
fraction of storage capacity used for storing the actual data).  For STAIR
codes, $E$ can be calculated by:
\begin{equation}\label{equ:effciency}
E = \frac{r \cdot (n-m) -s}{r \cdot n} \times 100\%.
\end{equation}
Note that the storage efficiency of Reed-Solomon codes can be obtained from
Equation~(\ref{equ:effciency}) by setting $s=0$, while that of an SD code with
a given $s$ \cite{Plank13b,Plank13c} can be directly computed via
Equation~(\ref{equ:effciency}).

Let $MTTDL_{arr}$ be the MTTDL of a storage array. Suppose that
$MTTDL_{arr}$ is exponentially distributed.  Then the MTTDL of the whole
storage system (denoted by $MTTDL_{sys}$) can be calculated by:
\begin{equation}\label{equ:mttdl-sys}
    MTTDL_{sys} = \frac{MTTDL_{arr}}{N_{arr}}.
\end{equation}

\begin{figure}[t]
\centering
\includegraphics[width=2.3in]{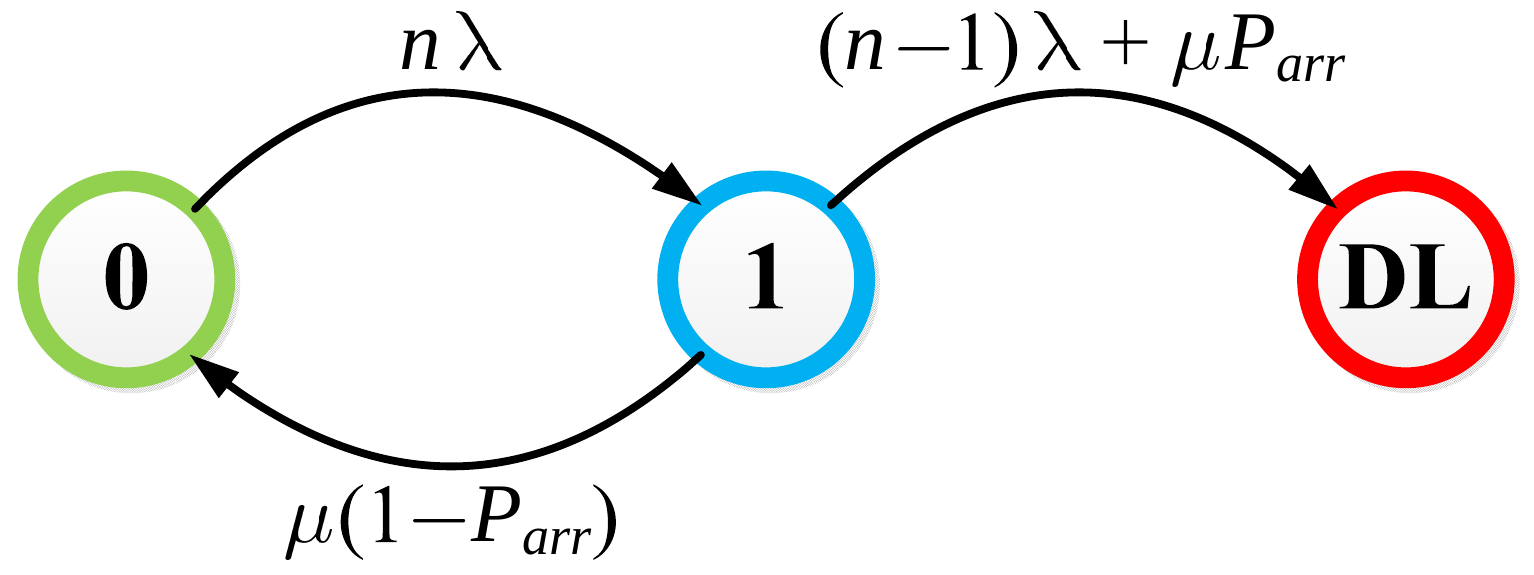}
\caption{Markov model for a storage array with $m=1$: State~0 means no device
failure; State~1 means one device failure; and State~DL means data loss.}
\label{fig:Markov}
\end{figure}

We first derive $MTTDL_{arr}$ as in the work \cite{Dholakia08}.
To simplify our analysis, we only consider the most practical case where
$m=1$.  When a storage array experiences a device failure, it enters
\emph{critical mode}, in which either an additional device failure or an
unrecoverable sector failure in a non-failed device can lead to data loss. For
device failures, suppose that they are independent and exponentially
distributed with parameter $\lambda$, where $1/\lambda$ is the mean time to
device failure; for sector failures, suppose that the probability that a
storage array in critical mode encounters unrecoverable sector failures in
non-failed devices is $P_{arr}$.  In addition, suppose that the rebuild time
in critical mode is exponentially distributed with parameter $\mu$, where
$1/\mu$ is the mean time to rebuild.  
Figure~\ref{fig:Markov} depicts the corresponding Markov model
\cite{Dholakia08}, where State~0 means no device failure, State~1 means one
device failure, and State~DL means data loss.  In this Markov model, we do not
consider the scenario where a storage array in State~0 encounters a sector
failure, by assuming that the storage array can recover the sector failure in
a very short time ($\ll 1/\mu$) and is highly unlikely to encounter another
device or sector failure that may lead to data loss.  An explicit expression
of $MTTDL_{arr}$ deduced based on this Markov model can be derived as
follows \cite{Dholakia08}:
\begin{equation}\label{equ:mttdl-m1}
MTTDL_{arr} = \frac{(2n-1)\lambda + \mu}{n\lambda[(n-1)\lambda +
	\mu P_{arr}]}.
\end{equation}

We next derive $P_{arr}$.  Recall that each stripe is independently encoded
in a storage array (see \S\ref{sec:problem}).  Let $P_{str}$ be the
probability that a stripe in critical mode encounters unrecoverable sector
failures in non-failed chunks.  Since the number of stripes in a storage array
is $\left\lfloor \frac{C}{S\cdot r} \right\rfloor$, where $S$ is the sector
size in bytes (typically 512~bytes), we have
\begin{equation}\label{equ:p-array}
    P_{arr} = 1 - (1-P_{str})^ {\left\lfloor\frac{C}{S\cdot r}\right\rfloor}
	\approx \left\lfloor\frac{C}{S\cdot r}\right\rfloor \cdot P_{str}.
\end{equation}

Finally, we discuss how to derive $P_{str}$.  In critical mode, there are
$n-m$ non-failed chunks in a stripe.  Suppose that each non-failed chunk
independently suffers from sector failures.   Let $P_{chk(i)}$ (where $0\leq i
\leq r$) be the probability that a non-failed chunk encounters $i$ sector
failures.   For the STAIR code with a given $\mb{e}$, we compute
$P_{str}$ as a function of $P_{chk(i)}$'s by enumerating all cases of
sector failures.  For example, if $\mb{e} = (s)$ (where $s\ge 1$), then
$P_{str}$ can be computed by the complement of the probability that all
$n-m$ non-failed chunks have no sector failure or exactly one non-failed
chunk has one up to $s$ sector failures.
Appendix~\ref{sec:appendix-expressions} describes the explicit expressions of
$P_{str}$ for some specific configurations of $\mb{e}$ considered in our
analysis.  For comparisons, Appendix~\ref{sec:appendix-expressions} also
describes the explicit expressions of $P_{str}$ for Reed-Solomon codes
and SD codes.  Note that the values of $P_{chk(i)}$'s are determined by the
sector failure model, which we describe below.




\subsubsection{Sector Failure Models}

Let $P_{sec}$ be the probability of a sector failure, and $P_{bit}$ be
the probability of an unrecoverable bit error.  Suppose that bit errors
are independent.  Then $P_{sec}$ can be estimated by:
\begin{equation}\label{equ:p-sector}
    P_{sec} = 1 - (1-{P}_{bit})^{S \times 8} \approx (S \times 8) \cdot {P}_{bit}.
\end{equation}

We now consider two models for sector failures \cite{Dholakia08}: the
independent model and the correlated model.  We fix $P_{sec}$ in both
models, so both models see the same expected number of sector failures in the
whole array.  Intuitively, in the independent model, we assume that sector
failures occur independently, so sector failures tend to be scattered across
different chunks within a stripe.  In the correlated model, we assume that
sector failures come in bursts, according to the previous field studies
\cite{Bairavasundaram07,Schroeder10}. Thus, sector failures tend to appear
together in one of the chunks within a stripe.  We derive $P_{chk(i)}$ (where
$0\leq i \leq r$) for each model as follows.

In the independent model, $P_{chk(i)}$ (where $0 \leq i \leq r$) is calculated
by:
\begin{equation}\label{equ:p-chunk}
    P_{chk(i)} = {r \choose i} \cdot P_{sec}^i \cdot (1-P_{sec})^{r-i}.
\end{equation}

In the correlated model, 
let $B$ be the average length (in number of sectors) of a sector failure
burst.  While the burst length may vary across different bursts, it is shown
that the average length $B$ is close to one sector (e.g., 
$B=1.0291$ \cite{Dholakia08}).  To
simplify our analysis, we assume that the burst length is at most $r$ sectors
in all cases, and that a burst spans one chunk only (i.e., it does not span
across two chunks).   We further assume that sector failure bursts are
independent of each other.  Let $b_i$ be the fraction of sector failure bursts
of length $i$ (where $1\leq i\leq r$) in a storage array (note that
$\sum_{i=1}^{r} {b_i}=1$). Then, we have:
%
\begin{equation}\label{equ:avg-B}
    B=\sum_{i=1}^{r} {i \times b_i}.
\end{equation}

Note that the probability that a sector is the beginning of a sector failure
burst is given by $P_{sec} \cdot \frac{1}{B}$.  Moreover, $P_{chk(0)}$ is equal to
the probability that each of the $r$ sectors in a chunk is not the beginning
of a sector failure burst.  Thus, we have:
\begin{equation}\label{equ:p-chunk-0}
   P_{chk(0)} = (1-\frac{P_{sec}}{B})^{r} \approx 1 - r \cdot \frac{P_{sec}}{B}.
\end{equation}
In other words, the probability that a chunk encounters at least one sector
failure is:
\begin{equation}\label{equ:p-chunk-nonzero}
    P_{chk(1)} + P_{chk(2)} + \cdots + P_{chk(r)} = 1 - P_{chk(0)} \approx r \cdot \frac{P_{sec}}{B}.
\end{equation}
We can compute $P_{chk(i)}$ (where $1 \leq i \leq r$) as:
\begin{equation}\label{equ:p-chunk-nonzero}
    P_{chk(i)} = b_i \cdot \left(r \cdot \frac{P_{sec}}{B}\right).
\end{equation}

\subsection{Numerical Results}


We examine the system reliability $MTTDL_{sys}$ of STAIR codes and compare it
with those of Reed-Solomon codes and SD codes.  We follow the storage array
configurations in the work \cite{Dholakia08}.  We consider a storage system
that stores $U=10$PB of user data using SATA disk drives with parameters
$C=300$GB, $S=512$~bytes, $1/\lambda=500,000$~hours, and $1/\mu=17.8$~hours.
In each storage array, we fix $n=8$, $r=16$, and $m=1$. We consider different
values of $s$ (note that $s=0$ corresponds to Reed-Solomon codes).  For a
given $s$, to store 10PB of user data, we set the number $N_{arr}$ of storage
arrays as follows:
\begin{center}
\begin{tabular}{c|ccccccc}
   \hline
   $s$ & 0 & 1 & 2 & 3 & 4 & 5 & 6  \\
   \hline
   $N_{arr}$ & 4994 & 5039 & 5085 & 5131 & 5179 & 5227 & 5276 \\
   \hline
   \hline
   $s$ & 7 & 8 & 9 & 10 & 11 & 12 & \\
   \hline
   $N_{arr}$ & 5327 & 5378 & 5430 & 5483 & 5538 & 5593 & \\
   \hline
\end{tabular}
\end{center}

For the probability $P_{bit}$ of an unrecoverable bit error in SATA disk
drives, we pick the range $[10^{-14},10^{-10}]$ to cover the data sheet value
$10^{-14}$ considered by Dholakia {\em et al.} \cite{Dholakia08} and the
empirical values that are much higher than stated in data sheets
\cite{Iliadis08b}.  We investigate how $P_{bit}$ affects the system
reliability.


\subsubsection{Independent Sector Failures}

\begin{figure}[t]
\centering
\subfigure[STAIR/SD codes with $s \leq 2$ and RS codes]{
    \label{fig:reliability_independent_different_p_a}
    \includegraphics[width=2.55in]{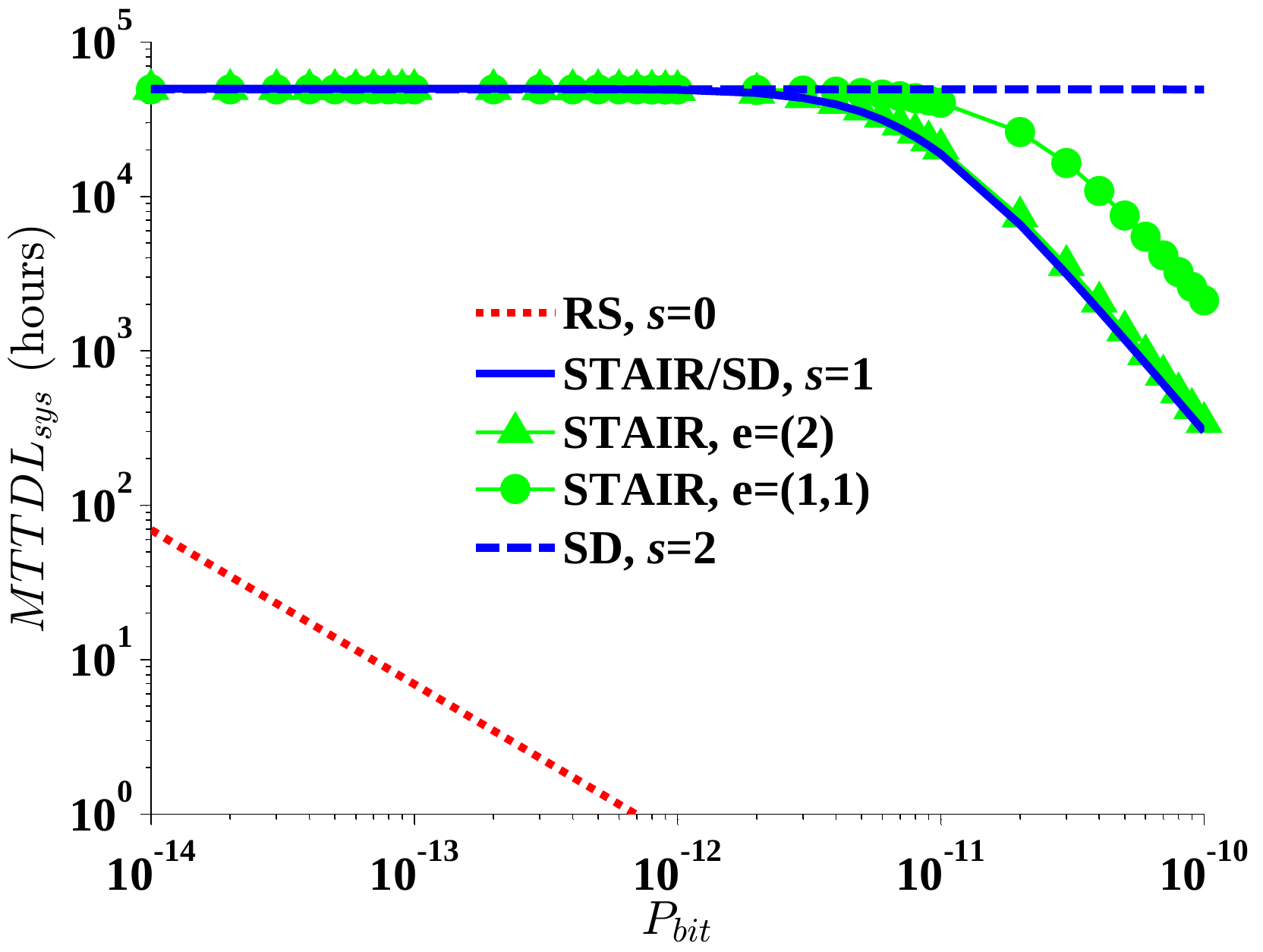}
}
\hspace{0.1in}
\subfigure[STAIR codes with $s = 3$]{
    \label{fig:reliability_independent_different_p_b}
    \includegraphics[width=2.55in]{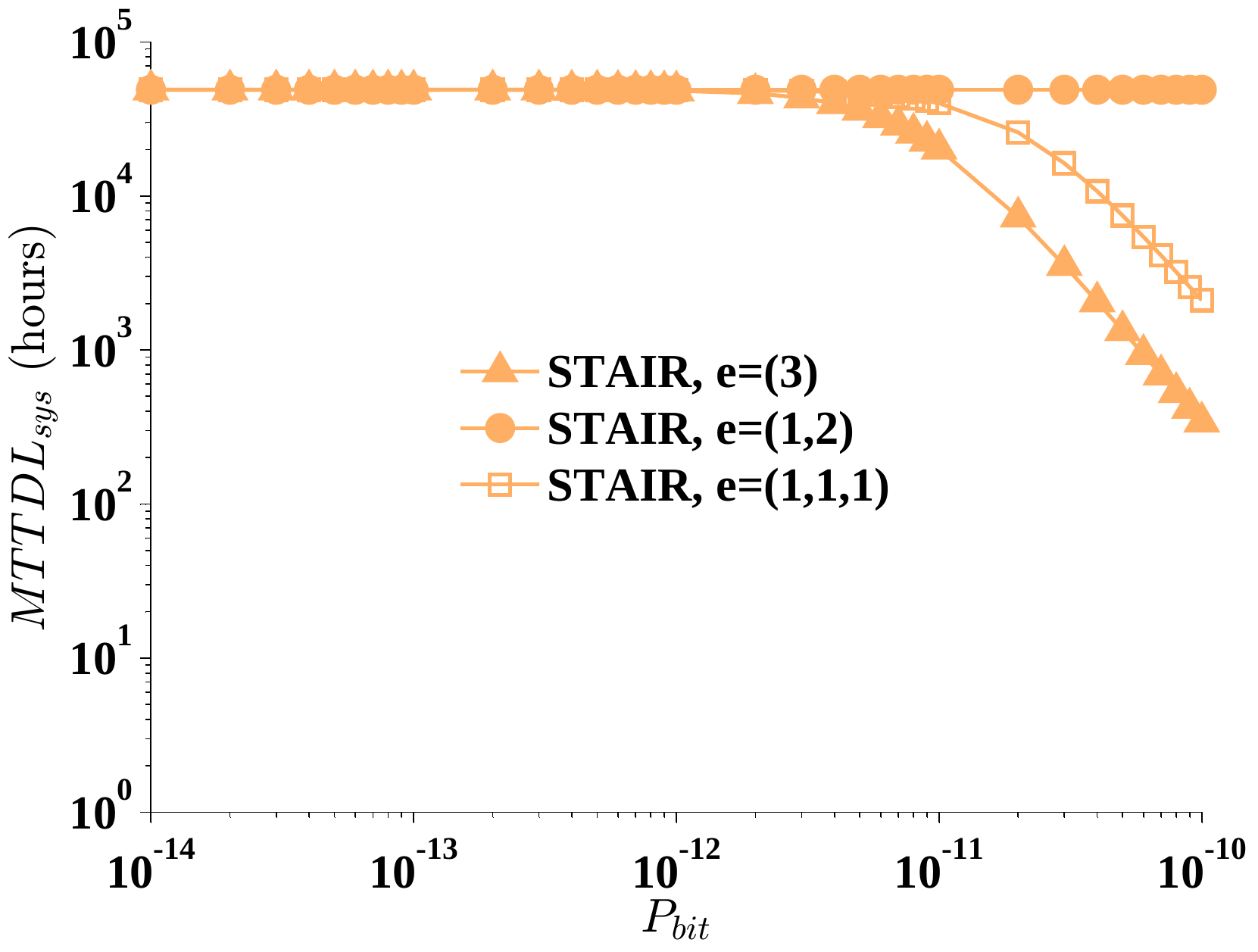}
}
\caption{$MTTDL_{sys}$ results of STAIR codes, SD codes, and Reed-Solomon (RS)
codes for different $P_{bit}$'s in the independent sector failure model.}
\label{fig:reliability_independent_different_p}
\end{figure}

We first consider the case of independent sector failures.
Figure~\ref{fig:reliability_independent_different_p} depicts $MTTDL_{sys}$
results of different erasure codes versus $P_{bit}$.  
From Figure~\ref{fig:reliability_independent_different_p}(a), we observe that
both the STAIR code and SD code with $s=1$ achieve much higher reliability than
Reed-Solomon codes, for example, by more than two orders of magnitude at
$P_{bit} = 10^{-14}$.  As $P_{bit}$ increases, the reliability of Reed-Solomon
codes follows a power-law decrease, while those of the STAIR code and SD code
with $s=1$ remain almost unchanged. The reason is that both STAIR codes and SD
codes can protect against the data loss due to an additional 
sector failure with an additional parity sector.  As $P_{bit}$ further
increases (beyond the order of $10^{-12}$), a storage array is more likely to
encounter more than one sector failure in critical mode,
and eventually has data loss before the rebuild finishes.  Thus, the
$MTTDL_{sys}$'s of both STAIR codes and SD codes drop (following
a power-law decrease).  Note that the decreasing trend of $MTTDL_{sys}$
observed here is similar to that observed by Dholakia {\em et al.}
\cite{Dholakia08}.



To improve the system reliability of both STAIR codes and SD codes, we choose
a higher value of $s$. For SD codes, if we choose $s=2$, its $MTTDL_{sys}$
remains almost unchanged over all $P_{bit}$'s we consider 
(see Figure~\ref{fig:reliability_independent_different_p}(a)); for STAIR
codes, we need to switch to $\mb{e}=(1,2)$ (i.e., $s=3$) to keep $MTTDL_{sys}$
unchanged (see Figure~\ref{fig:reliability_independent_different_p}(b)).  
Compared to SD codes, STAIR codes incur slightly higher storage
overhead (by storing one more parity sector per stripe) to achieve the same
reliability.  On the other hand, STAIR codes achieve much higher encoding
performance as observed in Figures~\ref{fig:encoding_speed} and
\ref{fig:encoding_stripe}.
		
Figure~\ref{fig:reliability_independent_different_p}(b) shows the
$MTTDL_{sys}$ results of STAIR codes for different configurations of $\mb{e}$,
all of which correspond to $s=3$.  Interestingly, $\mb{e}=(1,2)$ shows the
highest reliability. It has higher reliability than $\mb{e}=(3)$ because it
can protect against the sector failures that span more than one chunk
(horizontally), and it has higher reliability than $\mb{e}=(1,1,1)$ since it
can protect against more than one sector failure in a chunk (vertically).  

\subsubsection{Correlated Sector Failures}

We now consider the case of correlated sector failures, in which sector
failure bursts can occur.  Schroeder {\em et al.} \cite{Schroeder10} discover
that the length distribution of sector failure bursts can be fitted with a
pair of parameters: $({b}_1,\alpha)$, where ${b}_1$ is the fraction of sector
failure bursts of length one, and $\alpha$ ($>0$) is the tail index of a
Pareto distribution that best fits the distribution of burst length greater
than one.  A smaller $\alpha$ means a more heavy-tailed Pareto distribution.
Typically, ${b}_1$ often falls into the range between 0.9 and 0.99, and
$\alpha$ often falls into the range between 1 and 2
\cite[Table~1]{Schroeder10}. 

\begin{figure}[t]
\centering
\subfigure[STAIR codes ($s\le 2$), SD codes ($s \leq 2$), and RS codes]{
\label{fig:reliability_correlated_different_p_a}
\includegraphics[width=2.55in]{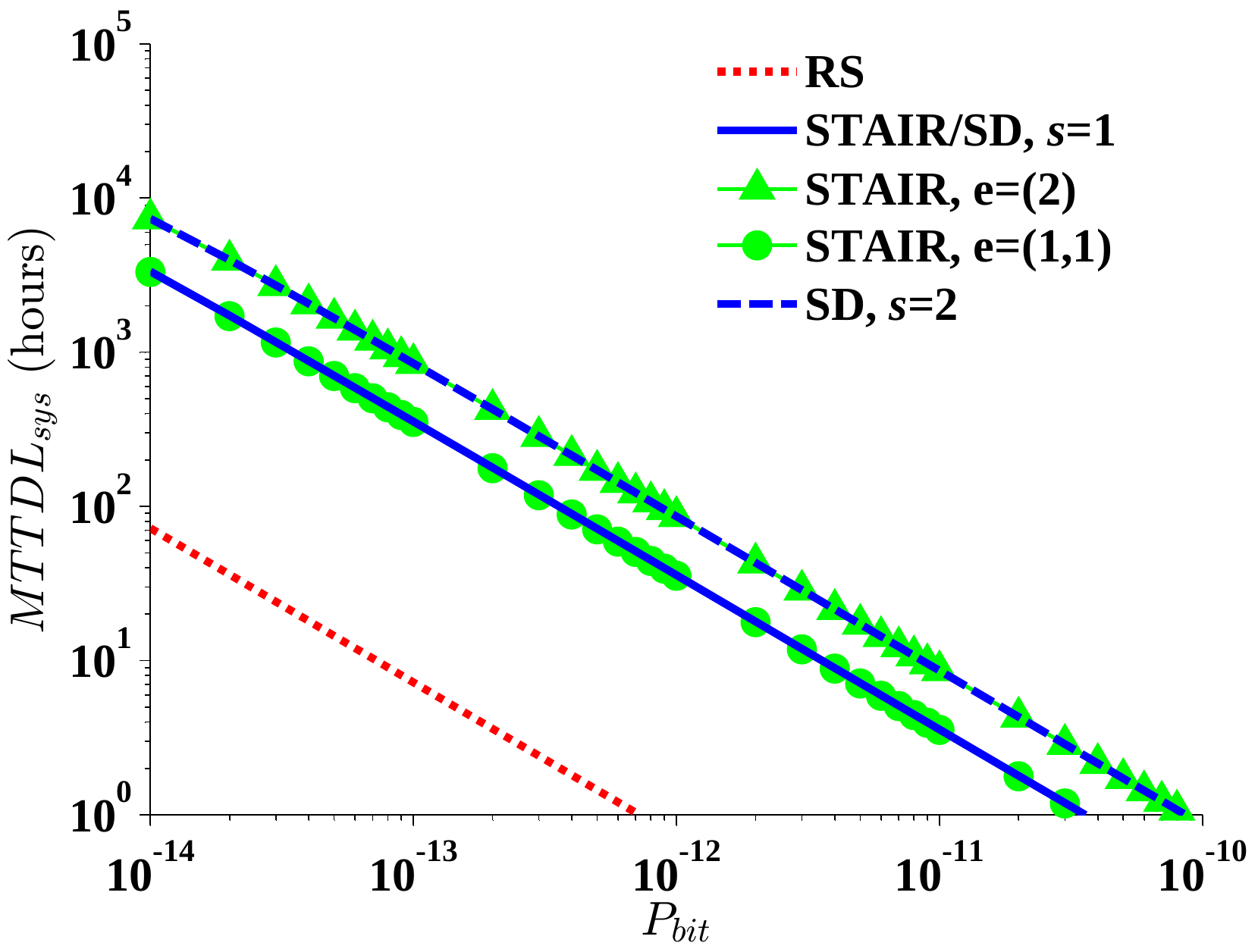}
}
\hspace{0.1in}
\subfigure[STAIR codes ($s = 3$) and SD codes ($s \leq 3$)]{
\label{fig:reliability_correlated_different_p_b}
\includegraphics[width=2.55in]{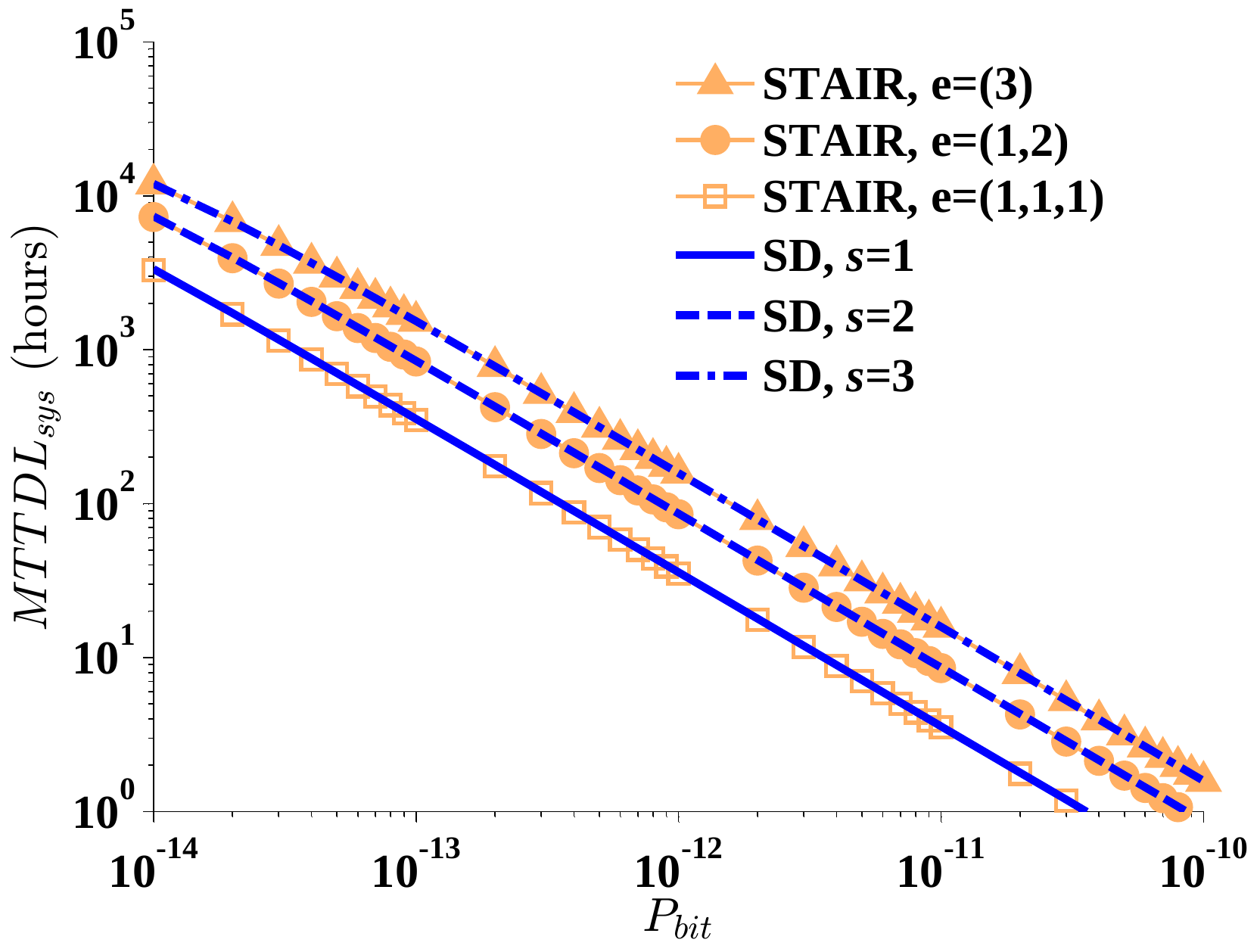}
}
\caption{$MTTDL_{sys}$ results of STAIR codes, SD codes, and Reed-Solomon (RS)
codes for different $P_{bit}$'s in the correlated sector failure
model with ${b}_1=0.98$ and $\alpha=1.79$.}
\label{fig:reliability_correlated_different_p}
\end{figure}

Figure~\ref{fig:reliability_correlated_different_p} first shows the impact
of $P_{bit}$ on $MTTDL_{sys}$.  Here, we consider a specific length
distribution of sector failure bursts where ${b}_1=0.98$ and $\alpha=1.79$
based on the ``D-2'' drive model in the work \cite{Schroeder10}.  
The reliability characteristics in the correlated sector failure model are very
different from those in the independent sector failure model.  From 
Figure~\ref{fig:reliability_correlated_different_p}(a), we observe that as
$P_{bit}$ increases, STAIR codes, SD codes, and Reed-Solomon codes show a
power-law decrease in reliability.  Nevertheless, both STAIR codes and SD
codes are more reliable than Reed-Solomon codes.  For example, when $P_{bit} =
10^{-14}$, both the STAIR code and SD code with $s=1$ achieve higher reliability
than Reed-Solomon codes by more than one order of magnitude.  In addition, 
from Figure~\ref{fig:reliability_correlated_different_p}(b), we observe that
the STAIR code with $\mb{e}=(e_0,e_1,\cdots,e_{m'-1})$ has almost the same
reliability as the SD code with $s=e_{m'-1}$ (e.g., see the $MTTDL_{sys}$'s of
the STAIR code with $\mb{e}=(1,2)$ and the SD code with $s=2$).  Also, among
all configurations of $\mb{e}$'s under the same $s$, the STAIR code with
$\mb{e}=(s)$ provides the highest reliability, which is almost the same as
that of the SD code with the same $s$ (e.g., see the $MTTDL_{sys}$'s of the
STAIR code with $\mb{e}=(3)$ and the SD code with $s=3$).  The reason is that in
our configuration of the correlated sector failure model, most sector failures
come as a burst that appears in one chunk.  Thus, the STAIR code with
$\mb{e}=(s)$ effectively protects against a sector burst of length $s$ in any
chunk, and has the same protection as the SD code with the same $s$. 

\begin{figure}[t]
\centering
\subfigure[Cumulative distribution functions (CDFs) of the length of sector failure bursts for different $({b}_1,\alpha)$ values]{
    \label{fig:reliability_correlated_large_s_probability}
    \includegraphics[width=5.5in]{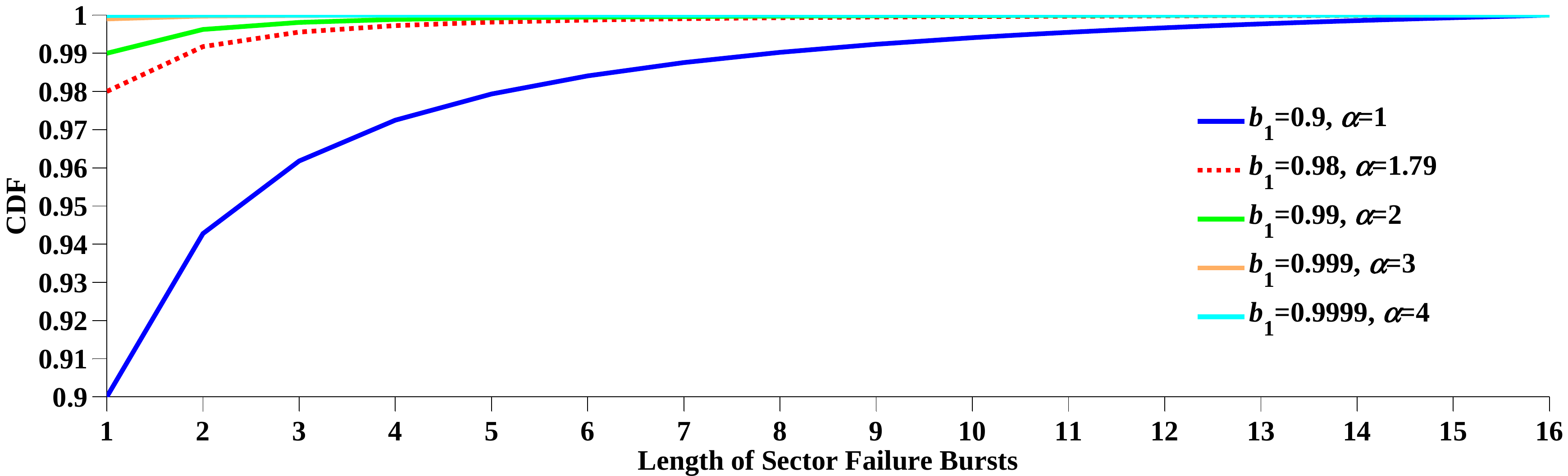}
}
\hspace{1in}
\subfigure[$MTTDL_{sys}$ results of STAIR codes with $\mb{e}=(s)$ and $\mb{e}=(1,s-1)$ for different $s$'s under different $({b}_1,\alpha)$ values]{
    \label{fig:reliability_correlated_large_s_value}
    \includegraphics[width=5.5in]{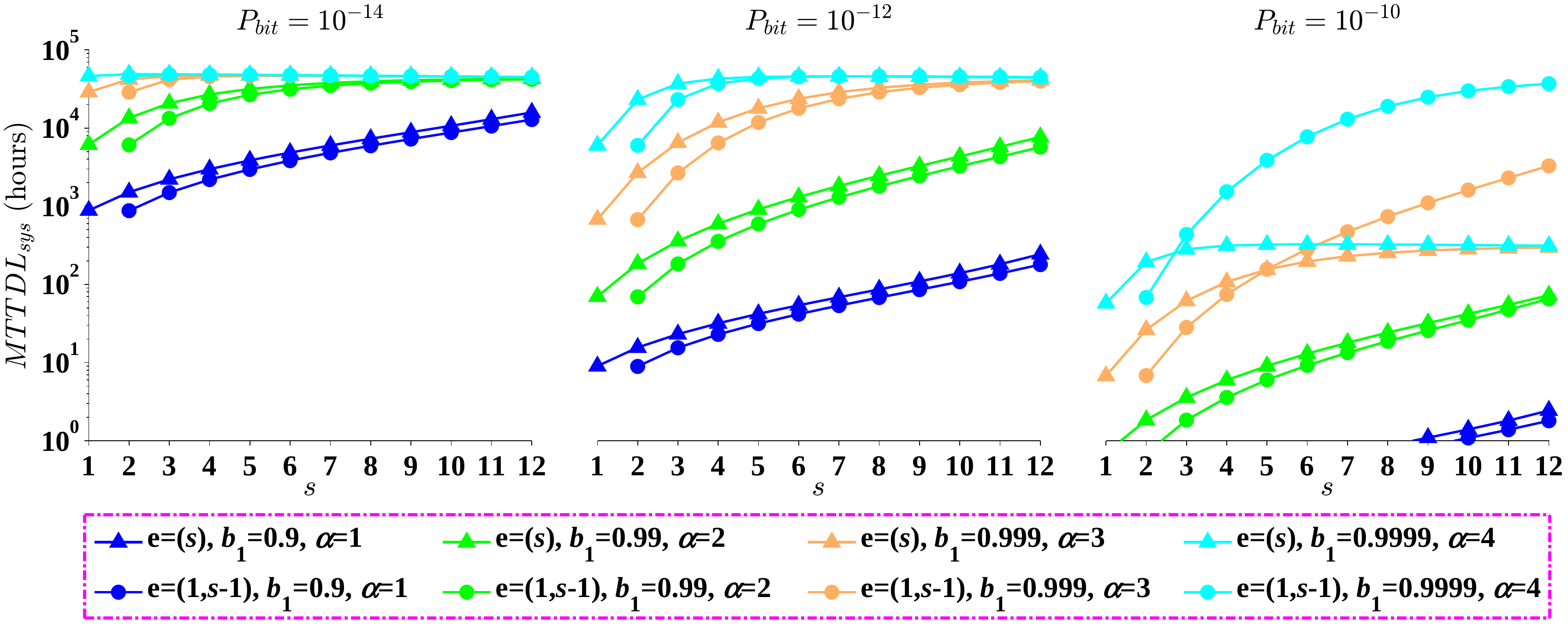}
}
\caption{$MTTDL_{sys}$ results of STAIR codes with $\mb{e}=(s)$ and
	$\mb{e}=(1,s-1)$ for different $s$'s in the correlated sector failure
	model with different $({b}_1,\alpha)$ values when $n=8$, $r=16$, and $m=1$.}
\label{fig:reliability_correlated_large_s}
\end{figure}

Figure~\ref{fig:reliability_correlated_large_s} next shows the impact of the
length distribution of sector failure bursts on $MTTDL_{sys}$.  Here, we only
consider STAIR codes, which can protect against sector failure bursts
of any length.  Figure~\ref{fig:reliability_correlated_large_s}(a) depicts
the burst length distribution for different pairs of $({b}_1,\alpha)$ that we
consider.  Smaller values of $b_1$ and $\alpha$ imply that the length of a
sector failure burst is more likely to be greater than one, or in other words,
sector failures are {\em more bursty}.
Figure~\ref{fig:reliability_correlated_large_s}(b) presents the $MTTDL_{sys}$
results of STAIR codes with $\mb{e}=(s)$ and $\mb{e}=(1,s-1)$ for different
values of $s$ under different pairs of $({b}_1,\alpha)$.  We observe that for
more bursty sector failures (e.g., ${b}_1=0.9$ and $\alpha=1$), the STAIR code
with $\mb{e}=(s)$ (for $s\ge 2$) achieves significantly higher reliability than
the STAIR code with $\mb{e}=(1,s-1)$.  In particular, as $s$ increases, the
reliability of the STAIR code with $\mb{e}=(s)$ increases exponentially.
This demonstrates the significance of STAIR codes that support a wider range
of $s$. On the other hand, for less bursty sector failures (e.g.,
${b}_1=0.9999$ and $\alpha=4$), as $s$ increases, the reliability of the STAIR
code with $\mb{e}=(s)$ increases much more slowly, and in some cases, is even
lower than that with $\mb{e}=(1,s-1)$ (e.g., when $P_{bit}=10^{-10}$).  This
observation is consistent with that in the independent sector failure model,
in which sector failures are likely scattered across different chunks within a
stripe. 
	 


\section{Related Work}
\label{sec:related}

Erasure codes have been widely adopted to provide fault tolerance against
device failures in storage systems \cite{Plank13a}.  Classical erasure codes
include standard Reed-Solomon codes \cite{Reed60} and Cauchy Reed-Solomon
codes \cite{Blomer95}, both of which are MDS codes that provide general
constructions for all possible configuration parameters. They are usually
implemented as systematic codes for storage applications
\cite{Plank97,Plank05,Plank06}, and thus can be used to implement the
construction of STAIR codes.  In addition, Cauchy Reed-Solomon codes can be
further transformed into {\em array codes}, whose encoding computations purely
build on efficient XOR operations \cite{Plank06}.


In the past decades, many kinds of array codes have been proposed, including
MDS array codes (e.g.,
\cite{Blaum95,Blaum96,Corbett04,Blaum06,Huang05,Feng05a,Feng05b,Xu99a,Xu99b,Li11,Plank11}) and
non-MDS array codes (e.g., \cite{Hafner05,Hafner06,Li09}).
Array codes are often designed for specific
configuration parameters.  To avoid compromising the generality of STAIR
codes, we do not suggest to adopt array codes in the construction of STAIR
codes.  Moreover, recent work \cite{Plank13d} has shown that Galois Field
arithmetic can be implemented to be extremely fast (sometimes at cache
line speeds) using SIMD instructions in modern processors.

Sector failures are not explicitly considered in traditional erasure codes,
which focus on tolerating device-level failures.  To cope with sector
failures, ad hoc schemes are often considered. One scheme is \emph{scrubbing}
\cite{Schwarz04,Oprea10,Schroeder10}, which proactively scans all disks and
recovers any spotted sector failure using the underlying erasure codes.
Another scheme is \emph{intra-device redundancy}
\cite{Dholakia08,Dholakia11,Schroeder10}, in which contiguous sectors in each
device are grouped together to form a segment and are then encoded with
redundancy within the device.	
Our work targets a different objective and focuses on constructing an erasure
code that explicitly addresses sector failures.

To simultaneously tolerate device and sector failures with minimal redundancy,
SD codes \cite{Plank13b,Plank13c} (including the earlier PMDS codes
\cite{Blaum13a}, which are a subset of SD codes) have recently been proposed.
As stated in \S\ref{sec:intro}, SD codes are known only for limited
configurations and some of the known constructions rely on extensive searches.
A relaxation of the SD property has also been
recently addressed as a future work \cite{Plank13c}, which assumes that
each row has no more than a given number of sector failures.  It is important to
note that the relaxation of \cite{Plank13c} is different from ours, in which
we limit the maximum number of devices with sector failures and the maximum
number of sector failures that simultaneously occur in each such device.  It
turns out that our relaxation enables us to derive a general code
construction.

There are other similar kinds of erasure codes that have similar constructions
to STAIR codes but serve for different purposes. Blaum \emph{et al.}
\cite{Blaum12} have constructed a family of nested codes that define the
number of tolerable sector failures in each row for an SSD array in which
sector failures appear as worn-out blocks.
However, unlike STAIR codes, such nested codes do not consider sector failure
bursts \cite{Bairavasundaram07,Schroeder10}. 
Another kind of erasure codes is the family of locally repairable codes (LRCs)
\cite{Huang12,Huang13,Sathiamoorthy13}, which focus on improving the recovery
performance of storage systems.  Pyramid codes \cite{Huang13} are designed for
small-scale device failures and have been implemented in archival storage
\cite{Wildani09}.
Huang {\em et al.}'s and Sathiamoorthy {\em et al.}'s LRCs
\cite{Huang12,Sathiamoorthy13} can be viewed as generalizations of Pyramid
codes and are recently adopted in commercial storage systems.
In particular, Huang {\em et al.}'s LRCs \cite{Huang12} achieve the same fault
tolerance property as PMDS codes \cite{Blaum13a}, and thus can also be
used as SD codes. However, the construction of Huang {\em et al.}'s LRCs
is limited to $m=1$ only.  To the best of our knowledge, STAIR codes are the
first general family of erasure codes that can efficiently tolerate both
device and sector failures.

\section{Conclusions}
\label{sec:conclusions}

We present STAIR codes, a general family of erasure codes that can tolerate
simultaneous device and sector failures in a space-efficient manner.
STAIR codes can be constructed for tolerating any numbers of device and sector
failures subject to a pre-specified sector failure coverage.
The special construction of STAIR codes also makes efficient
encoding/decoding possible through parity reuse.  Compared to the
recently proposed SD codes \cite{Blaum13a,Plank13b,Plank13c}, STAIR codes not
only support a much wider range of configuration parameters, but also achieve
higher encoding/decoding speed based on our experiments.



The source code of STAIR codes is available at
{\bf http://ansrlab.cse.cuhk.edu.hk/software/stair}.


\section*{APPENDIX}

\appendix
\section{Proof of Homomorphic Property}
\label{sec:appendix-Homomorphic}

We formally prove the homomorphic property described in
\S\ref{subsec:homomorphism}.  We state the following theorem.

\begin{theorem}\label{th:homomorphic}
In the construction of the canonical stripe of STAIR codes, the encoding of
each chunk in the column direction via $\mc{C}_{col}$ is homomorphic, such
that each augmented row in the canonical stripe is a codeword of
$\mc{C}_{row}$.
\end{theorem}

\noindent
{\bf Proof:} We prove by matrix operations.  We define the matrices
$\mb{D} = [d_{i,j}]_{r\times(n-m)}$, $\mb{P} = [p_{i,k}]_{r\times m}$, and
$\mb{P}' = [p'_{i,l}]_{r\times m'}$.  Also, we define the generator matrices
$\mb{G}_{row}$ and $\mb{G}_{col}$ for the codes $\mc{C}_{row}$ and
$\mc{C}_{col}$, respectively, as:
\begin{eqnarray*}
\mb{G}_{row} & = & \left(\mb{I}_{(n-m)\times(n-m)} \mid
		\mb{A}_{(n-m)\times(m+m')} \right), \\
\mb{G}_{col} & = & \left(\mb{I}_{r\times r}
		\mid \mb{B}_{r\times e_{m'-1}}\right),
\end{eqnarray*}
where $\mb{I}$ is an identity matrix, and $\mb{A}$ and $\mb{B}$ are the
sub-matrices that form the parity symbols.  The upper $r$ rows of the stripe
can be expressed as follows:
\begin{eqnarray*}
\left(\mb{D} \mid \mb{P} \mid \mb{P}'\right) & = & \mb{D} \cdot \mb{G}_{row}.
\end{eqnarray*}
The lower $e_{m'-1}$ augmented rows are expressed as follows:
\begin{eqnarray*}
\left(\left(\mb{D} \mid \mb{P} \mid \mb{P}'\right)^T \cdot \mb{B}\right)^T
& = & \mb{B}^T \cdot \left( \mb{D} \cdot \mb{G}_{row} \right) \\
& = & \left(\mb{B}^T \cdot \mb{D} \right) \cdot \mb{G}_{row}
\end{eqnarray*}
We can see that each of the lower $e_{m'-1}$ rows can be calculated using
the generator matrix $\mb{G}_{row}$, and hence is a codeword of $\mc{C}_{row}$.
$\hfill\Box$

\section{Explicit Expressions of $P_{str}$ for Various Erasure Codes}
\label{sec:appendix-expressions}

\subsection{Reed-Solomon Codes}

The explicit expression of $P_{str}$ for Reed-Solomon codes is as follows:
\begin{equation}
    P_{str} = 1 - P_{chk(0)}^{n-m}.
\end{equation}

\subsection{STAIR Codes}

Explicit expressions of $P_{str}$ for some STAIR codes with special $\mb{e}$'s are as follows:

\begin{enumerate}
  \item For a STAIR code with $\mb{e}=(s)$ for $s \geq 1$,
\begin{equation}
    P_{str} = 1 - P_{chk(0)}^{n-m} -  {n-m \choose 1} \cdot \sum\limits_{i=1}^{s} P_{chk(i)} \cdot P_{chk(0)}^{n-m-1}.
\end{equation}
  \item For a STAIR code with $\mb{e}=(1,s-1)$ for $s \geq 2$,
\begin{equation}
  \begin{split}
    P_{str} = & 1 - P_{chk(0)}^{n-m} -  {n-m \choose 1} \cdot \sum\limits_{i=1}^{s-1} P_{chk(i)} \cdot P_{chk(0)}^{n-m-1} -  \\
                & {n-m \choose 2} \cdot P_{chk(1)}^2 \cdot P_{chk(0)}^{n-m-2} - {n-m \choose 1} \cdot {n-m-1 \choose 1} \cdot \sum\limits_{i=2}^{s-1} P_{chk(i)} \cdot P_{chk(1)}  \cdot P_{chk(0)}^{n-m-2}.
  \end{split}
\end{equation}
  \item For a STAIR code with $\mb{e}=(2,s-2)$ for $s \geq 4$,
\begin{equation}
  \begin{split}
    P_{str} = & 1 - P_{chk(0)}^{n-m} -  {n-m \choose 1} \cdot \sum\limits_{i=1}^{s-2} P_{chk(i)} \cdot P_{chk(0)}^{n-m-1} - \\
                & {n-m \choose 2} \cdot P_{chk(1)}^2 \cdot P_{chk(0)}^{n-m-2} - {n-m \choose 1} \cdot {n-m-1 \choose 1} \cdot \sum\limits_{i=2}^{s-2} P_{chk(i)} \cdot P_{chk(1)} \cdot P_{chk(0)}^{n-m-2} - \\
                & {n-m \choose 2} \cdot P_{chk(2)}^2 \cdot P_{chk(0)}^{n-m-2} - {n-m \choose 1} \cdot {n-m-1 \choose 1} \cdot \sum\limits_{i=3}^{s-2} P_{chk(i)} \cdot P_{chk(2)} \cdot P_{chk(0)}^{n-m-2}.
  \end{split}
\end{equation}
  \item For a STAIR code with $\mb{e}=(1,1,s-2)$ for $s \geq 3$,
\begin{equation}
  \begin{split}
    P_{str} = & 1 - P_{chk(0)}^{n-m} -  {n-m \choose 1} \cdot \sum\limits_{i=1}^{s-2} P_{chk(i)} \cdot P_{chk(0)}^{n-m-1} - \\
                & {n-m \choose 2} \cdot P_{chk(1)}^2 \cdot P_{chk(0)}^{n-m-2} -  {n-m \choose 1} \cdot {n-m-1 \choose 1} \cdot \sum\limits_{i=2}^{s-2} P_{chk(i)} \cdot P_{chk(1)} \cdot P_{chk(0)}^{n-m-2} - \\
                & {n-m \choose 3} \cdot P_{chk(1)}^3 \cdot P_{chk(0)}^{n-m-3} -  {n-m \choose 2} \cdot {n-m-2 \choose 1} \cdot \sum\limits_{i=2}^{s-2} P_{chk(i)} \cdot P_{chk(1)}^2 \cdot P_{chk(0)}^{n-m-3}.  \\
  \end{split}
\end{equation}
  \item For a STAIR code with $\mb{e}=(\overset{s}{\overbrace{1,1,\cdots,1}})$ for $s \geq 1$,
\begin{equation}
    P_{str} = 1 - \sum\limits_{i=0}^{s} \left({n-m \choose i} \cdot
			P_{chk(1)}^{i} \cdot P_{chk(0)}^{n-m-i} \right).
\end{equation}
\end{enumerate}

\subsection{SD Codes}

Explicit expressions of $P_{str}$ for SD codes with $s \leq 3$ \cite{Plank13b,Plank13c} are as follows:

\begin{enumerate}
  \item For an SD code with $s=1$,
\begin{equation}
    P_{str} = 1 - P_{chk(0)}^{n-m} -  {n-m \choose 1} \cdot P_{chk(1)} \cdot P_{chk(0)}^{n-m-1}.
\end{equation}
  \item For an SD code with $s=2$,
\begin{equation}
    P_{str} = 1 - P_{chk(0)}^{n-m} -  {n-m \choose 1} \cdot \sum\limits_{i=1}^{2} P_{chk(i)} \cdot P_{chk(0)}^{n-m-1} - {n-m \choose 2} \cdot P_{chk(1)}^2 \cdot P_{chk(0)}^{n-m-2}.
\end{equation}
  \item For an SD code with $s=3$,
\begin{equation}
  \begin{split}
    P_{str} = & 1 - P_{chk(0)}^{n-m} -  {n-m \choose 1} \cdot \sum\limits_{i=1}^{3} P_{chk(i)} \cdot P_{chk(0)}^{n-m-1} - \\
                & {n-m \choose 2} \cdot P_{chk(1)}^2 \cdot P_{chk(0)}^{n-m-2} - {n-m \choose 1} \cdot {n-m-1 \choose 1} \cdot P_{chk(2)} \cdot P_{chk(1)} \cdot P_{chk(0)}^{n-m-2} - \\
                & {n-m \choose 3} \cdot P_{chk(1)}^{3} \cdot P_{chk(0)}^{n-m-3}.
  \end{split}
\end{equation}
\end{enumerate}


\bibliographystyle{abbrv}
\bibliography{reference}

\end{document}